\pdfoutput=1
\documentclass[aps,prd,reprint,superscriptaddress]{revtex4-1}

\usepackage{graphicx}% Include figure files
\usepackage{lineno}
\usepackage{amsmath}
\usepackage{amssymb}
\usepackage{amsthm}
\usepackage{booktabs} % for much better looking tables

\begin{document}

%\linenumbers

%\preprint{}

\title{Cosmogenic neutron production at Daya Bay}


%\author{Some Author}
%\thanks{}
%\affiliation{Some Institution}


%%% Generation information for this file
%%% Wed Mar  7 09:49:39 2018
%%% /Users/djaffe/Documents/Reactor/work/trunk/NuWa-trunk/dybgaudi/Documentation/AuthorList/Neutron2016
%%% ../AuthGen.py dyb_collabotion_list_20160701.xls 20160701 
\newcommand{\ECUST}{\affiliation{Institute of Modern Physics, East China University of Science and Technology, Shanghai}}
\newcommand{\IHEP}{\affiliation{Institute~of~High~Energy~Physics, Beijing}}
\newcommand{\Wisconsin}{\affiliation{University~of~Wisconsin, Madison, Wisconsin 53706}}
\newcommand{\Yale}{\affiliation{Wright~Laboratory and Department~of~Physics, Yale~University, New~Haven, Connecticut 06520}} %%%% 20170227 add Wright Lab
\newcommand{\BNL}{\affiliation{Brookhaven~National~Laboratory, Upton, New York 11973}}
\newcommand{\NTU}{\affiliation{Department of Physics, National~Taiwan~University, Taipei}}
\newcommand{\NUU}{\affiliation{National~United~University, Miao-Li}}
\newcommand{\Dubna}{\affiliation{Joint~Institute~for~Nuclear~Research, Dubna, Moscow~Region}}
\newcommand{\CalTech}{\affiliation{California~Institute~of~Technology, Pasadena, California 91125}}
\newcommand{\CUHK}{\affiliation{Chinese~University~of~Hong~Kong, Hong~Kong}}
\newcommand{\NCTU}{\affiliation{Institute~of~Physics, National~Chiao-Tung~University, Hsinchu}}
\newcommand{\NJU}{\affiliation{Nanjing~University, Nanjing}}
\newcommand{\TsingHua}{\affiliation{Department~of~Engineering~Physics, Tsinghua~University, Beijing}}
\newcommand{\SZU}{\affiliation{Shenzhen~University, Shenzhen}}
\newcommand{\NCEPU}{\affiliation{North~China~Electric~Power~University, Beijing}}
\newcommand{\Siena}{\affiliation{Siena~College, Loudonville, New York  12211}}
\newcommand{\IIT}{\affiliation{Department of Physics, Illinois~Institute~of~Technology, Chicago, Illinois  60616}}
\newcommand{\LBNL}{\affiliation{Lawrence~Berkeley~National~Laboratory, Berkeley, California 94720}}
\newcommand{\UIUC}{\affiliation{Department of Physics, University~of~Illinois~at~Urbana-Champaign, Urbana, Illinois 61801}}
\newcommand{\RPI}{\affiliation{Department~of~Physics, Applied~Physics, and~Astronomy, Rensselaer~Polytechnic~Institute, Troy, New~York  12180}}
\newcommand{\SJTU}{\affiliation{Department of Physics and Astronomy, Shanghai Jiao Tong University, Shanghai Laboratory for Particle Physics and Cosmology, Shanghai}}
\newcommand{\BNU}{\affiliation{Beijing~Normal~University, Beijing}}
\newcommand{\WM}{\affiliation{College~of~William~and~Mary, Williamsburg, Virginia  23187}}
\newcommand{\Princeton}{\affiliation{Joseph Henry Laboratories, Princeton~University, Princeton, New~Jersey 08544}}
\newcommand{\VirginiaTech}{\affiliation{Center for Neutrino Physics, Virginia~Tech, Blacksburg, Virginia  24061}}
\newcommand{\CIAE}{\affiliation{China~Institute~of~Atomic~Energy, Beijing}}
\newcommand{\SDU}{\affiliation{Shandong~University, Jinan}}
\newcommand{\NanKai}{\affiliation{School of Physics, Nankai~University, Tianjin}}
\newcommand{\UC}{\affiliation{Department of Physics, University~of~Cincinnati, Cincinnati, Ohio 45221}}
\newcommand{\DGUT}{\affiliation{Dongguan~University~of~Technology, Dongguan}}
\newcommand{\XJTU}{\affiliation{Department of Nuclear Science and Technology, School of Energy and Power Engineering, Xi'an Jiaotong University, Xi'an}}%%updated 20161215 {Xi'an Jiaotong University, Xi'an}}
\newcommand{\UCB}{\affiliation{Department of Physics, University~of~California, Berkeley, California  94720}}
\newcommand{\HKU}{\affiliation{Department of Physics, The~University~of~Hong~Kong, Pokfulam, Hong~Kong}}
\newcommand{\UH}{\affiliation{Department of Physics, University~of~Houston, Houston, Texas  77204}}
\newcommand{\Charles}{\affiliation{Charles~University, Faculty~of~Mathematics~and~Physics, Prague}} %updated20140819
\newcommand{\USTC}{\affiliation{University~of~Science~and~Technology~of~China, Hefei}}
\newcommand{\TempleUniversity}{\affiliation{Department~of~Physics, College~of~Science~and~Technology, Temple~University, Philadelphia, Pennsylvania  19122}}
\newcommand{\CUC}{\affiliation{Instituto de F\'isica, Pontificia Universidad Cat\'olica de Chile, Santiago}} % update 20140911 use r23215 for Spanish
\newcommand{\CGNPG}{\affiliation{China General Nuclear Power Group, Shenzhen}}% update 20170729 add city. updated 20140724 China~Guangdong~Nuclear~Power~Group, Shenzhen}}
\newcommand{\NUDT}{\affiliation{College of Electronic Science and Engineering, National University of Defense Technology, Changsha}} % added 20140111
\newcommand{\IowaState}{\affiliation{Iowa~State~University, Ames, Iowa  50011}}
\newcommand{\ZSU}{\affiliation{Sun Yat-Sen (Zhongshan) University, Guangzhou}}
\newcommand{\CQU}{\affiliation{Chongqing University, Chongqing}} % add 20150417
\newcommand{\BCC}{\altaffiliation[Now at ]{Department of Chemistry and Chemical Technology, Bronx Community College, Bronx, New York  10453}} % add 20160311 for Sunej's 'now at'
\author{F.~P.~An}\ECUST
\author{A.~B.~Balantekin}\Wisconsin
\author{H.~R.~Band}\Yale
\author{M.~Bishai}\BNL
\author{S.~Blyth}\NTU\NUU
\author{D.~Cao}\NJU
\author{G.~F.~Cao}\IHEP
\author{J.~Cao}\IHEP
\author{Y.~L.~Chan}\CUHK
\author{J.~F.~Chang}\IHEP
% ...

\collaboration{The Daya Bay Collaboration}

\date{\today}

\begin{abstract}

Neutrons produced by cosmic ray muons are an important background for underground experiments studying neutrino oscillations, neutrinoless double beta decay, dark matter, and other rare-event signals.  A measurement of the neutron yield in the three different experimental halls of the Daya Bay Reactor Neutrino Experiment at varying depth is reported.  The neutron yield in Daya Bay's liquid scintillator is measured to be $Y_n=(10.26\pm 0.86)\times 10^{-5}$, $(10.22\pm 0.87)\times 10^{-5}$, and $(17.03\pm 1.22)\times 10^{-5}\mu^{-1}~\textnormal{g}^{-1}~\textnormal{cm}^2$ at depths of 250, 265, and 860~meters-water-equivalent.  These results are compared to other measurements and the simulated neutron yield in {\sc Fluka} and {\sc Geant4}.  A global fit including the Daya Bay measurements yields a power law coefficient of $0.77 \pm 0.03$ for the dependence of the neutron yield on muon energy.
\end{abstract}

%\keywords{neutron, antineutrino detector, neutron yield}

\maketitle

\section{Introduction}

Neutrons and other hadrons produced by cosmic ray muons are an important source of background for underground low-background experiments studying neutrino oscillations, double beta decay, dark matter and other rare events. There have been several studies of cosmogenic neutron production. Muon-induced neutron and isotope production has been studied with the CERN Super Proton Synchrotron (SPS) muon beam in 2000~\cite{Hagner:2000xb}. Studies on neutron and isotope yields in various materials in underground detectors have been performed by the INFN large-volume detector (LVD)~\cite{Aglietta:1999iw}, Borexino~\cite{Bellini:2013pxa}, KamLAND~\cite{Abe:2009aa}, and many others~\cite{Aglietta:1989xn,Hertenberger:1995ae,Bezrukov, Boehm:2000ru,Enikeev,Blyth:2015nha,Araujo:2008ze,Reichhart:2013xkd}.  This paper reports a measurement of the neutron production rate in liquid scintillator at three different values of average muon energy by the Daya Bay Reactor Neutrino Experiment, an underground low-background neutrino oscillation experiment.

\begin{figure}[htp]
\centering
\includegraphics[width=0.45\textwidth]{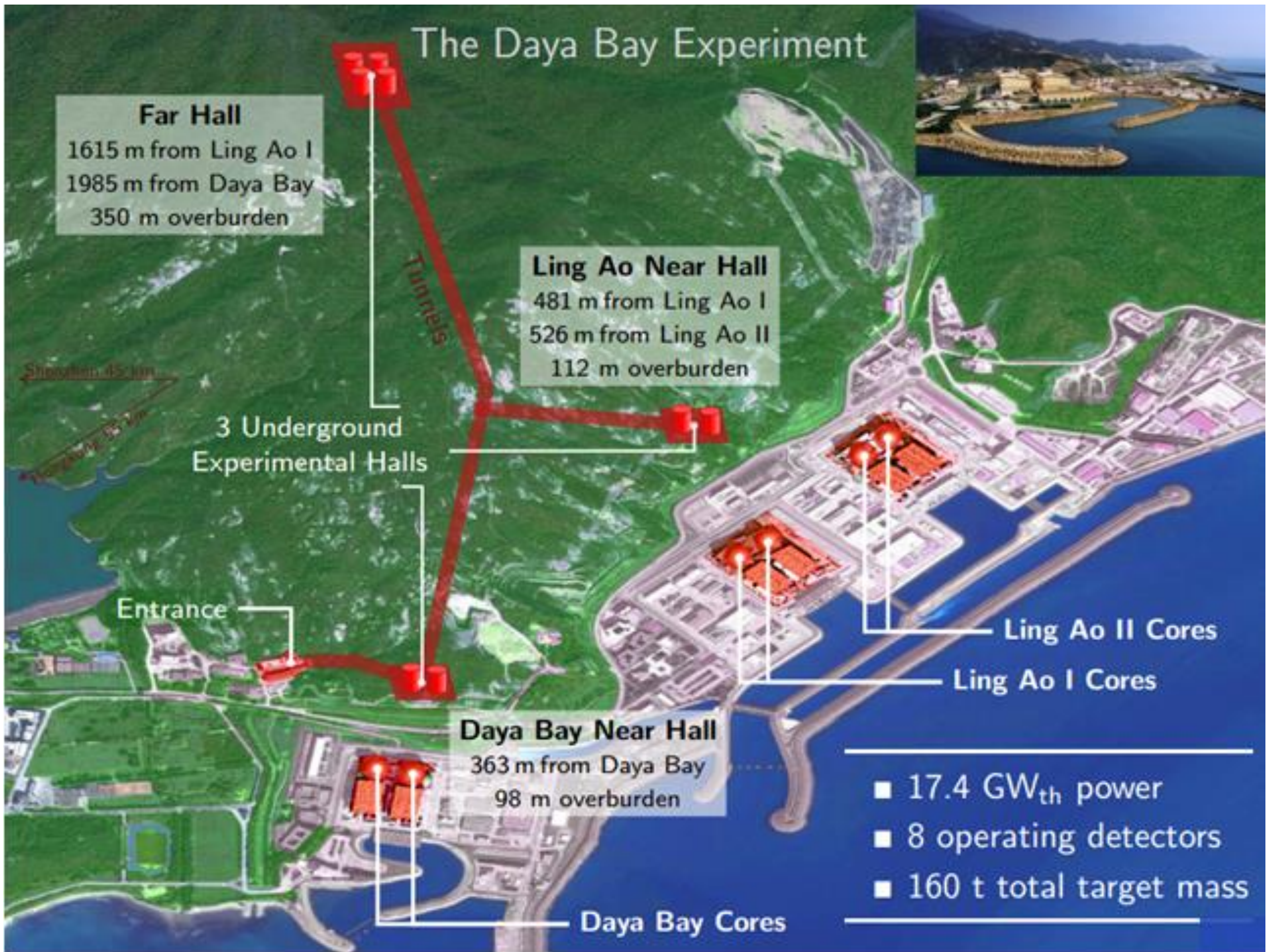}
\caption{\label{fig:Layout} 
A map of the layout of the Daya Bay Reactor Neutrino Experiment, including six reactor cores (Daya Bay, Ling Ao I, and Ling Ao II cores) and three experimental halls (two near and one far). The antineutrino detectors (ADs) are located in the underground experimental halls, with two ADs at the Daya Bay Near Hall (EH1), two at the Ling Ao Near Hall (EH2), and four at the Far Hall (EH3).
%A 3D representation (above) and 2D projection (below) of the layout of the Daya Bay Reactor Neutrino Experiment, including %six reactor cores (Daya Bay 1-2, Ling Ao 1-4) and three experimental halls (EH1-3). The antineutrino detectors (ADs) are %located in the underground experimental halls, with two ADs in EH1, two in EH2, and four in EH3.
}
\end{figure}

Daya Bay, located near the city of Shenzhen in the Guangdong province in China, is designed to study neutrino oscillations by measuring the survival probability of electron antineutrinos from nuclear reactors~\cite{Guo:2007ug}. 
Daya Bay has made increasingly precise measurements of $\sin^22\theta_{13}$~\cite{An:2012eh,An:2013uza,An:2013zwz,An:2015rpe,An:2016ses} and the effective neutrino mass-squared difference $|\Delta m^2_{\rm ee}|$~\cite{An:2013zwz,An:2015rpe,An:2016ses}.  Figure~\ref{fig:Layout} shows a diagram of the Daya Bay experimental site.  The Daya Bay Nuclear Power Plant complex consists of six reactors, producing 17.4~GW of total thermal power.  The experiment has three experimental halls (EHs), two halls near the reactors cores (EH1, EH2) and one hall far from the cores (EH3). 
Relative measurements in multiple detector sites are used to predominantly cancel reactor flux and spectral shape uncertainties.
In its full configuration, the experiment employs eight functionally-identical antineutrino detectors (ADs) to decrease detector-related errors, with two placed in each near hall and four in the far hall. The ADs are enclosed in water to shield against backgrounds and located underground to reduce the cosmic ray muon flux. Each site has redundant muon detectors to identify the residual muons.  The ADs are designed to identify neutron captures, providing the possibility to identify muon-induced neutrons.  The three EHs are at vertical depths of 250, 265, and 860~meters-water-equivalent (m.w.e.)~allowing for a measurement of the neutron yield at three different values of average muon energy within the same experiment.

\begin{figure*}[htp]
\centering
\includegraphics[width=0.4\textwidth]{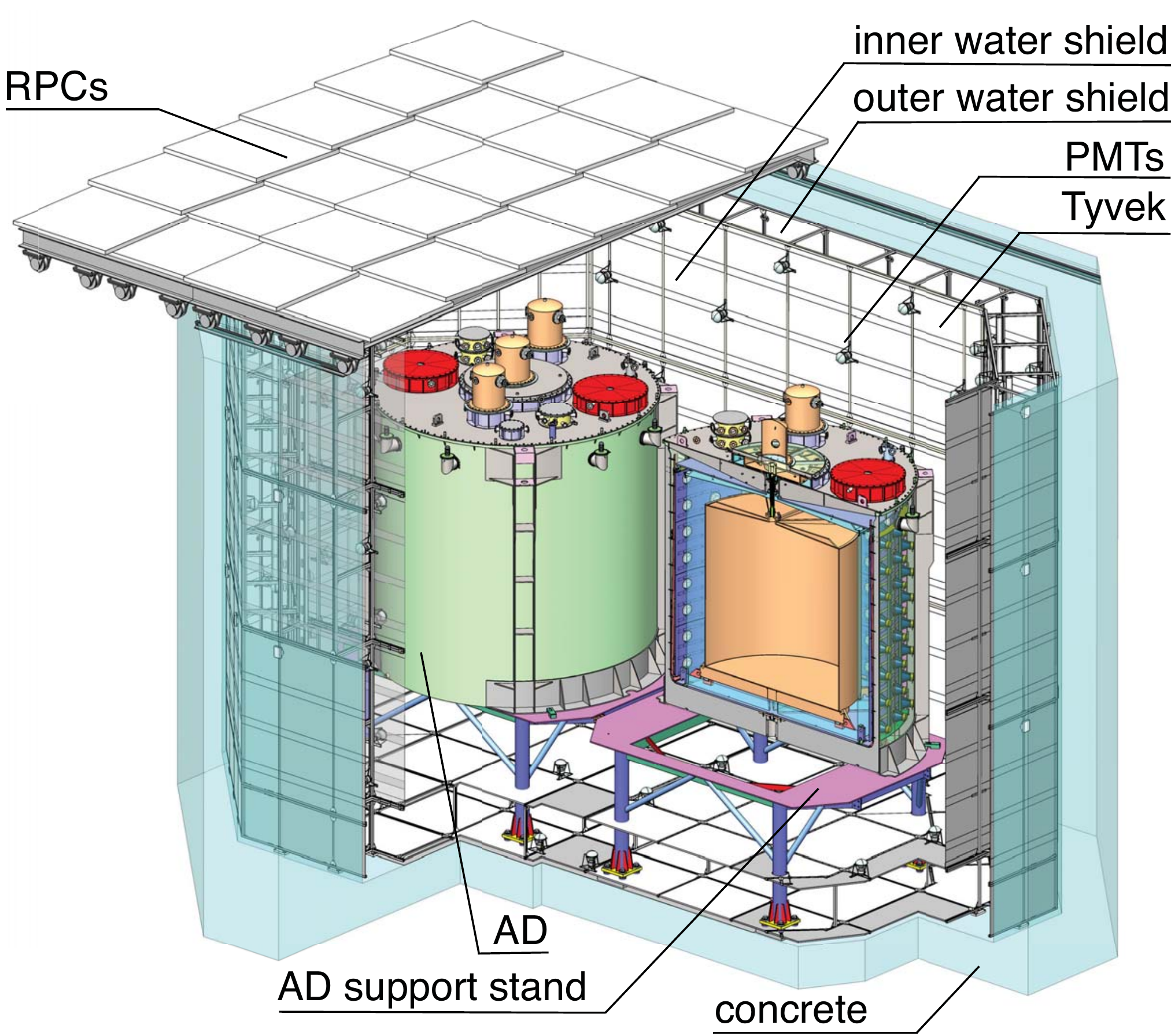}
\includegraphics[width=0.55\textwidth]{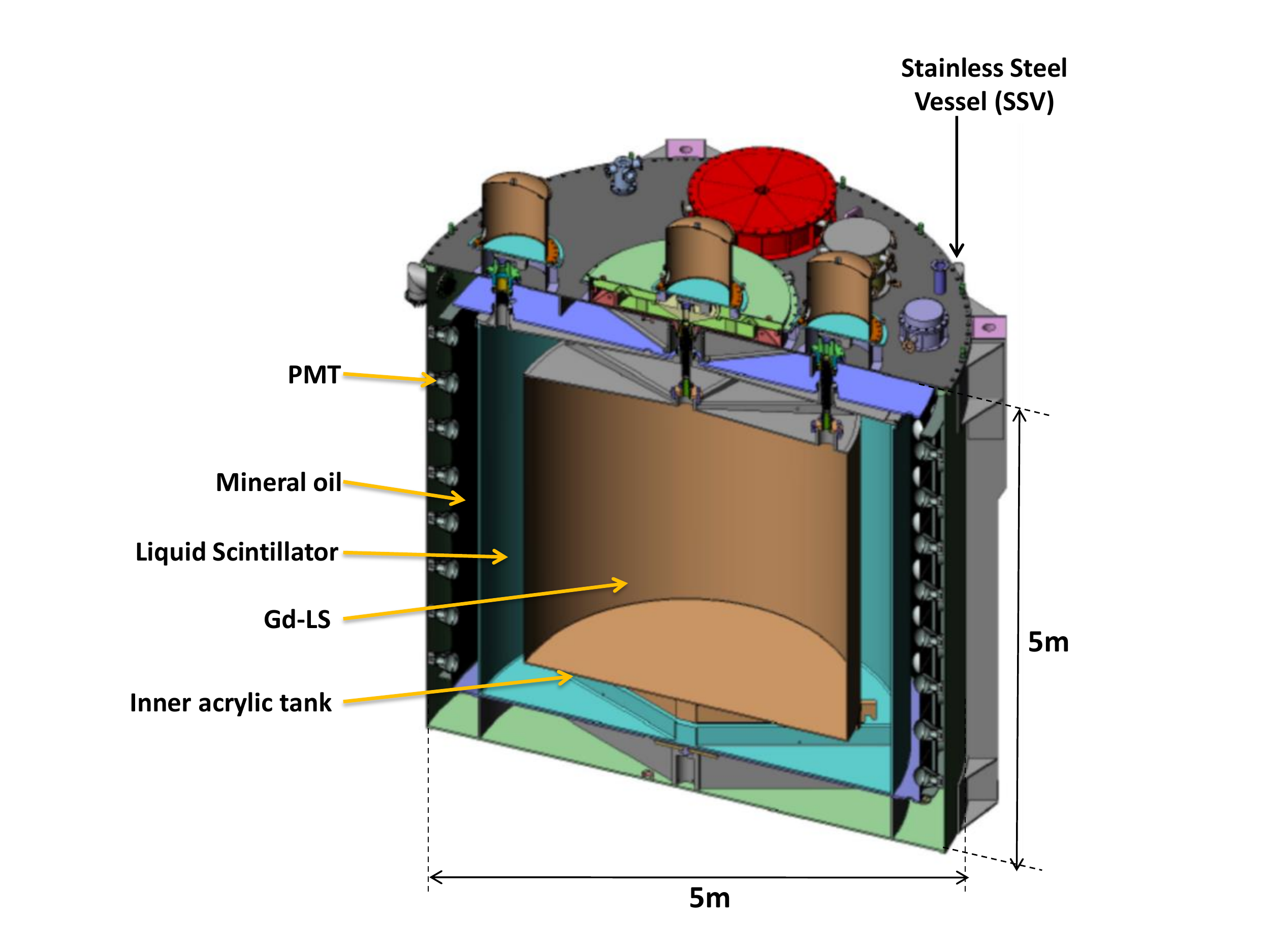}
\caption{\label{fig:Det} Left: Diagram of near site detectors. Right: Diagram of an AD.}
\end{figure*}

\section{Detectors}
 
\subsection{Antineutrino Detectors}
\label{sec:ads}
The Daya Bay ADs~\cite{DayaBay:2012aa,An:2015qga} detect antineutrino interactions via the inverse beta decay (IBD) reaction, $\bar{\nu}_e + p \to e^{+} + n $. The ADs, shown in Fig.~\ref{fig:Det}, consist of three concentric cylindrical regions separated by transparent acrylic vessels.  The central target region of each AD is 3~m in height and 3~m in diameter and is filled with 20~tons of liquid scintillator loaded with 0.1\% gadolinium by weight (GdLS).  The second layer, called the gamma catcher, is filled with liquid scintillator (LS) to detect neutron capture gamma rays that escape from the target region. The outer layer is a mineral oil buffer for additional shielding against radioactivity.  The stainless steel vessel surrounding the outermost layer is 5~m in height and 5~m in diameter.  Each detector contains 192 8-inch photomultiplier tubes (PMTs) distributed uniformly on the inside wall of the containment vessel. Reflectors on the top and bottom improve the light collection and uniformity. 

The IBD reaction is characterized by two time-correlated triggers, the prompt signal coming from the energy loss of the positron in the scintillator and its annihilation, and the delayed signal from the capture of the neutron.  The liquid scintillator is loaded with gadolinium (Gd) to increase the capture rate of thermal neutrons which suppresses backgrounds from accidental coincidences.
 The neutrons are preferentially captured on Gd, producing an 8-MeV gamma ray cascade, which is much higher than the energy range of most background radioactivity. 
%This is the characteristic neutron capture signal.  

\subsection{Muon Detectors}
The ADs are surrounded by a water shield, which also serves as a water Cherenkov counter providing 4$\pi$ veto coverage for muons traversing any AD. The water shield is covered by a cosmic-ray detector array made of resistive plate chambers (RPCs).  See Ref.~\cite{Dayabay:2014vka} for a detailed description of the muon system. 

Each AD is shielded from natural radioactive background and cosmogenic neutrons by at least 2.5~m of water in every direction. The water shield is instrumented with 288 (384) PMTs in the near (far) halls to detect muons via Cherenkov radiation.  It is optically separated into two individual water Cherenkov detectors, the inner water shield (IWS) and outer water shield (OWS), using a thin layer of diffusely reflecting Tyvek.  The OWS is 1~m thick on the sides and bottom of the water pool.  A water circulation and purification system is used to maintain water quality and detector performance~\cite{Wilhelmi:2014irz}.  Figure~\ref{fig:Det} shows a diagram of the near site muon detectors.

\begin{figure}[htp]
\centering
\includegraphics[width=0.45\textwidth]{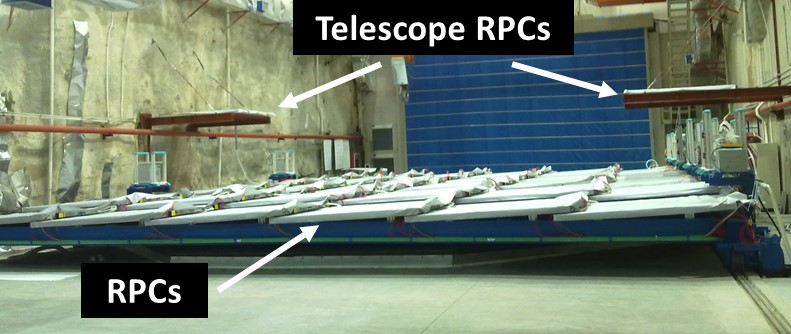}
\caption{\label{fig:TeleRPC} Photograph of EH1 showing the RPCs and telescope RPC system in position over the water pool.  The main RPCs are at floor level, and the two telescope RPCs extend from the wall on the left and right of the photograph.}
\end{figure}

 A system of RPC modules is installed above the water shield.  Each module has four RPC layers and has dimensions 2.17~m~$\times$~2.20~m~$\times$~0.08~m.  There are 54 modules in each near hall and 81 modules in the far hall.  In addition, two telescope RPC modules positioned 2~m above the main RPC systems in each hall are used for precise muon track reconstruction for a smaller portion of the solid angle to benchmark the muon simulation.  The position resolution of the RPCs is approximately 10~cm.  The RPCs and telescope RPCs are shown in Fig.~\ref{fig:TeleRPC}.

\section{Simulation}

The neutron yield is defined as the number of neutrons produced per muon, per path length of the muon through the material, per density of the material, which in this case is GdLS.  Neutrons captured on Gd following an identified AD muon are selected in the data, but corrections to this number are necessary to determine the yield.  For example, some neutrons produced by a muon in the GdLS will escape without being detected.  In this analysis, the corrections between the produced and detected number of neutrons are derived from a Monte Carlo (MC) simulation.  Muon track reconstruction is challenging in Daya Bay due to the detector size and the complicated geometry of the water shield.  Instead of determining the average path length of muons through the GdLS based on reconstructed tracks, which would introduce large uncertainties, the average muon path length as determined by simulation is used to calculate the yield.  Aspects of the simulation that are important for this analysis are described below.

\subsection{Muon Flux Simulation}
\label{sec:muonsim}
The sea level muon flux is well-described by Gaisser's formula ~\cite{Gaisser1,Eidelman:2004wy}. For Daya Bay, Gaisser's formula is modified to better describe the muon spectrum at low energies and large zenith angles~\cite{Guan}.
A digitized mountain profile is generated based on topographic maps of the Daya Bay Nuclear Power Plant region. 
The {\sc Music} code~\cite{Antonioli:1997qw,Kudryavtsev:2008qh} is used to propagate muons from the top of the mountain to the underground halls and uses the digitized mountain profile to calculate
the path length in rock. {\sc Music}'s standard rock properties ($Z=11$, $A = 22$, and $\rho = 2.65$~g/cm$^3$) are used in the simulation. 

\begin{table}[htp]
  \centering
  \caption{ \label{tab:underground} Underground muon simulation results. All values have been transformed into a detector-independent spherical geometry. The error in the simulated total flux is about 10\%.}
  \begin{tabular}{lccc}
    \hline
    \hline
Hall & \multicolumn{2}{c}{Overburden} & Muon flux \\
  & m & m.w.e. & Hz/m$^2$ \\
    \hline
EH1~~~ & 93 & 250 & 1.27 \\
EH2~~~ & 100 & 265 & 0.95 \\
EH3~~~ & 324 & 860 & 0.056 \\
\hline
\hline
\end{tabular}
\end{table}

\begin{figure}[htp]
\centering
    \includegraphics[width=0.45\textwidth]{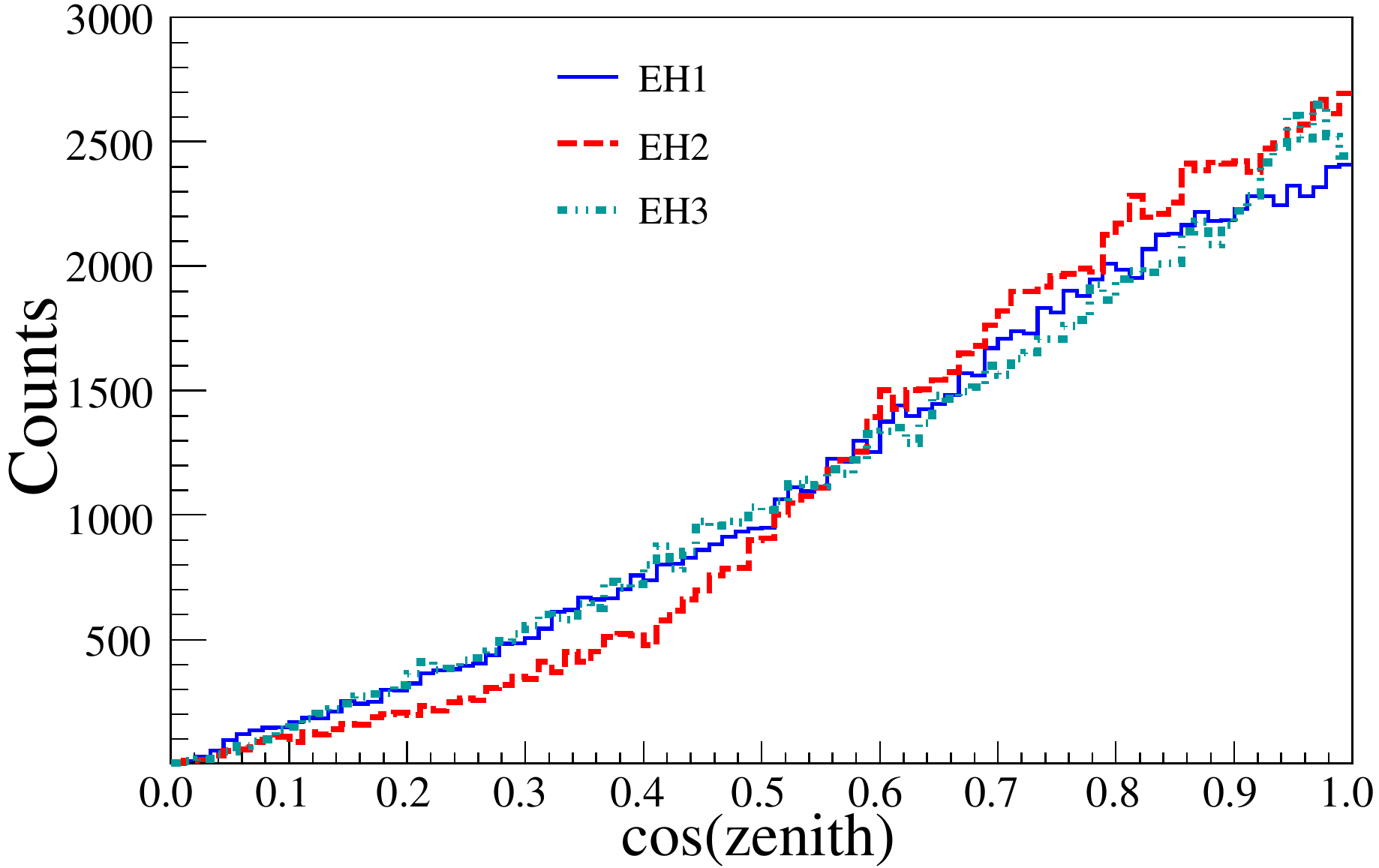}
    \includegraphics[width=0.45\textwidth]{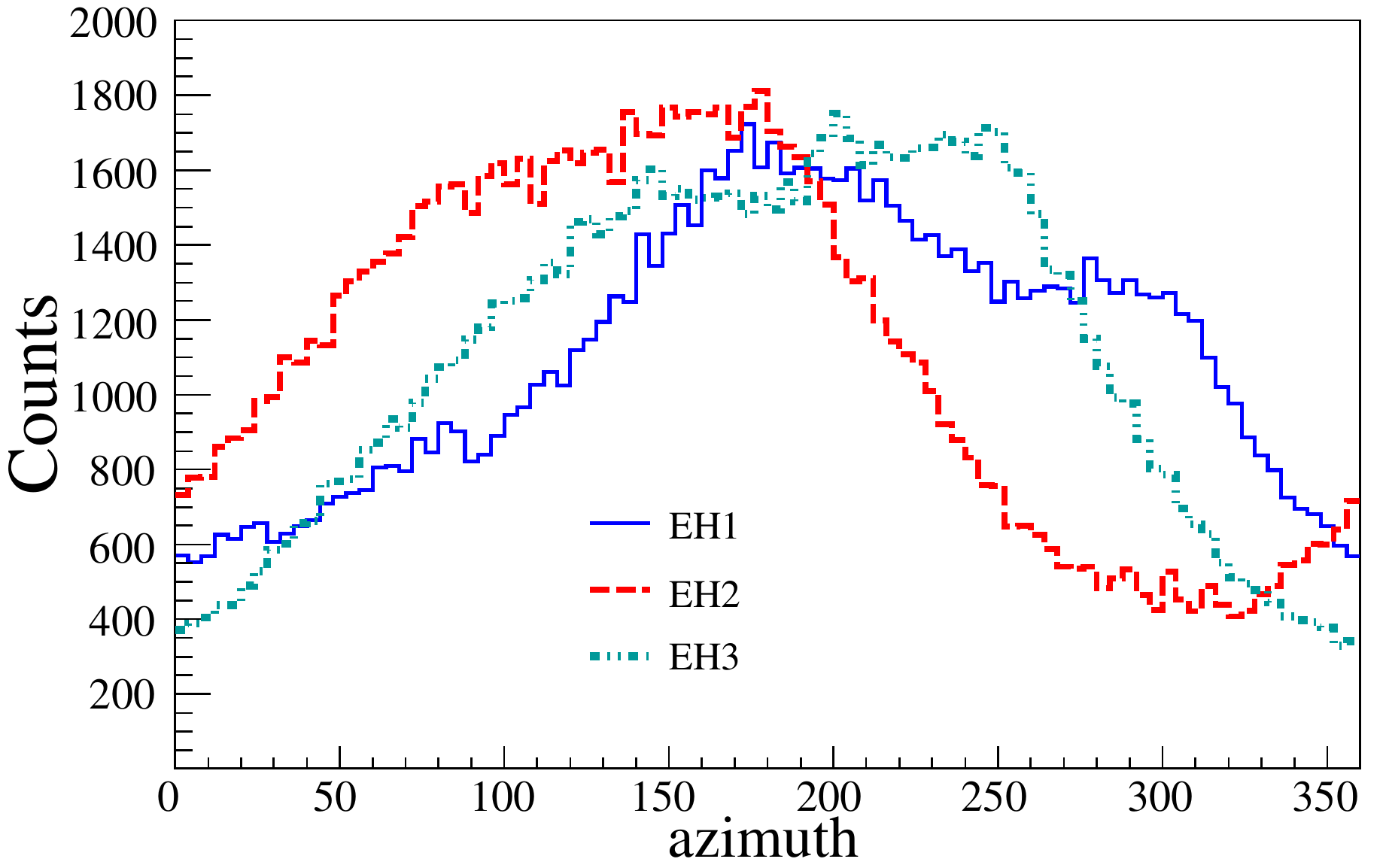}
    \caption{\label{fig:mu_sim_angles} Simulated trajectories of muons that reach the underground halls.  By definition, zenith is the angle from the vertical and azimuth is the horizontal compass angle from true North.  (A zenith angle of $0^{\circ}$ represents a downward-going muon, and an azimuthal angle of $0^{\circ}$ corresponds to a muon coming from the northern direction.)  Differences in angular distributions at each hall are due to the mountain profile.}
\end{figure}

\begin{figure}[htp]
\centering
    \includegraphics[width=0.49\textwidth]{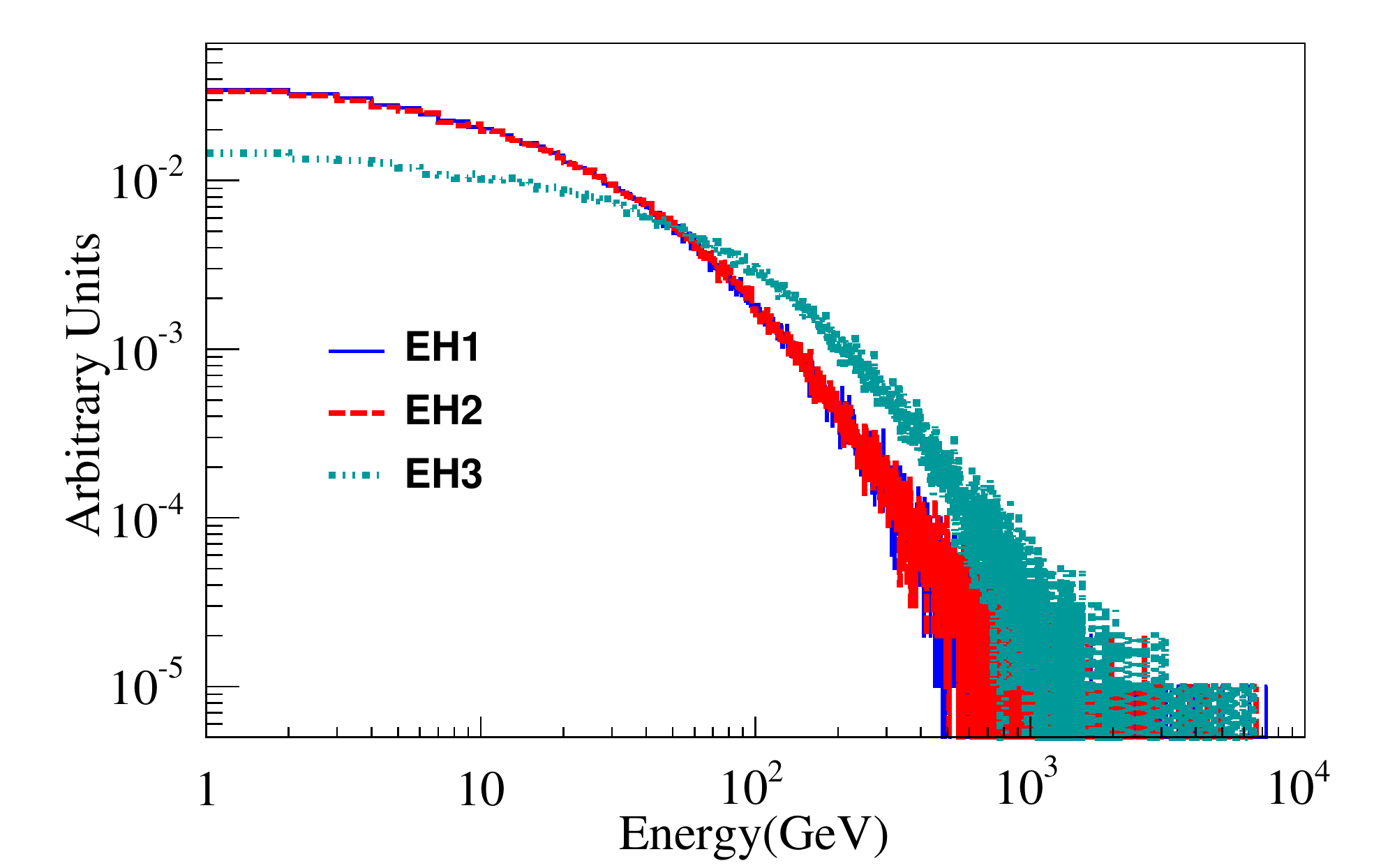}
    \caption{\label{fig:mu_sim_energy} Simulated energy spectra of muons that reach the underground halls.  The differences between EH1 and EH2 are too small to be visible at this scale.}
\end{figure}

Table~\ref{tab:underground} shows the underground muon flux for each hall from the {\sc Music} simulation. Figures~\ref{fig:mu_sim_angles} and \ref{fig:mu_sim_energy} show muon angle and energy distributions at each hall, respectively.
The zenith angle is defined as a muon's angle from the vertical, and azimuth is a muon's horizontal compass angle from true North.  Differences in angular distributions at each hall are due to the mountain profile.  The error in the total simulated muon flux is about 10\%, which includes the uncertainties in the mountain profile mapping, rock composition, density profiling, and {\sc Music} simulation.

\subsection{Detector Simulation}

The muons generated with {\sc Music} are used as the incident muon sample in the detector simulation to study neutron production by muons.  Approximately $4\times 10^{9}$, $2\times 10^{8}$, and $2\times 10^{9}$ muons are simulated in EH1, EH2, and EH3, respectively.

The Daya Bay detector MC simulation is based on {\sc Geant4}~\cite{Agostinelli:2002hh,Allison:2006ve}.
For the purpose of the neutron yield study, neutrons are also simulated in the Daya Bay detectors using {\sc Fluka}~\cite{fluka1,Ferrari:2005zk} as a cross-check of the default  {\sc Geant4} simulation. The features of both {\sc Geant4} and {\sc Fluka} relevant to the neutron yield analysis are described in this section.  Unless otherwise stated, the nominal {\sc Geant4}-based simulation is used in what follows.

For detailed studies of MC predictions for neutron production by cosmic ray muons for various depths and materials, and comparisons with experimental data, see Refs.~\cite{Wang:2001fq,Kudryavtsev:2003aua,Araujo:2004rv,Mei:2005gm,Empl:2014zea,Formaggio:2004ge}.

\subsubsection{{\sc Geant4}}
 {\sc Geant4} is a widely-used toolkit for simulating the passage of particles through matter~\cite{Agostinelli:2002hh,Allison:2006ve}.
{\sc Geant4} version 9.2p01 is used for this analysis. 
The physics list used in the simulation is QGS BIC, which applies the binary cascade
(BIC) model for hadronic interactions at lower energies (between 70~MeV and 9.9~GeV for protons and neutrons). For hadronic interactions at higher energy, a quark-gluon string (QGS) model is applied.  
{\sc Geant4}'s precompound model is used for hadronic interactions at the lowest energies (below 70~MeV) and as a nuclear de-excitation model within the higher-energy QGS model.
For neutron elastic and inelastic interactions below 20~MeV, a data-driven model is used (NEUTRONHP). 
For neutron capture on Gd, {\sc Geant4}'s default neutron capture library is modified to require energy conservation.

The full simulation is conducted without optical processes to increase the simulation speed.  Without optical photons, reconstruction algorithms based on PMT hits cannot be used to determine the muon's trajectory and energy deposition.  Therefore, the muon's path length and deposited energy are taken from the simulated path length and deposited energy. 
%The minimum path length for simulation is 1~mm for gammas and 100~$\mu$m for electrons and positrons.

%When a neutron interacts, {\sc Geant4} treats the incoming neutron and outgoing neutron as two distinct neutrons.  Due to this feature, special care is taken to avoid double-counting neutrons for the calculations described in Section~\ref{sect:nsel}.

\subsubsection{{\sc Fluka}}
{\sc Fluka} is another popular tool for simulations of particle transport and interactions with matter. {\sc Fluka} version 2011.2b is used for this analysis. 
The hadron-nucleon interaction model in {\sc Fluka} is based on resonance production and decay below a few
GeV and on the dual parton model at higher energies. For hadron-nucleus interactions, a nuclear interaction model called {\sc Peanut}~\cite{Ferrari:1993xr} is used.
For neutron interactions below 20~MeV, {\sc Fluka} uses its own neutron cross section library containing more than 250 different materials.

 The Daya Bay geometry is included in {\sc Fluka} at the same level of detail used in the {\sc Geant4} simulation, with the exception that PMTs are not included in the former. Similar to the {\sc Geant4} simulation, the muon's path length and deposited energy are taken from the simulated path length and deposited energy. 
%, and a threshold of 1~MeV for gammas is implemented to increase the simulation speed. Studies show that decreasing the threshold down to 0.1~MeV has no effect on the neutron yield.  
%The same considerations applied in {\sc Geant4} for neutron counting are applied for {\sc Fluka}.

\section{Neutron Yield}

\subsection{Analysis Strategy}

%When selecting IBD interactions of reactor antineutrinos, AD triggers occurring within a short time window after a detected muon are vetoed to remove cosmogenic background from the IBD sample. 
In this analysis, AD triggers following a detected muon are used to study neutrons produced by cosmic muons. 

The neutron yield $Y_n$ can be expressed as

\begin{equation}
\label{eqn:ny}
 Y_{n}=  \frac{N_{n}}{ N_{\mu} L_{\rm avg} \rho},
\end{equation}
where $N_{n}$ is the number of neutrons produced in association with $N_{\mu}$ muons traversing the GdLS target, $L_{\rm avg}$ is the average path length of muons in the GdLS from simulation, and $\rho = 0.86$~g/cm$^3$~\cite{An:2015qga} is the measured density of Daya Bay's GdLS.  The following sections explain the details of the selection of muons traversing the target and the selection of neutrons produced by these muons.

\subsection{Data set}
 The Daya Bay experiment began collecting data on 24 December 2011 with six ADs.  Two ADs were located in EH1, one AD in EH2 and three ADs in EH3.  In summer 2012, data taking was paused to install two new ADs, one in EH2 and one in EH3.  Operation restarted on 19 October 2012.  The results presented here are based on 404 days of data acquired with the full configuration of eight ADs and 217 days of data acquired with six ADs.

\subsection{Muon event selection}
\begin{figure}[htp]
\centering
\includegraphics[width=0.49\textwidth]{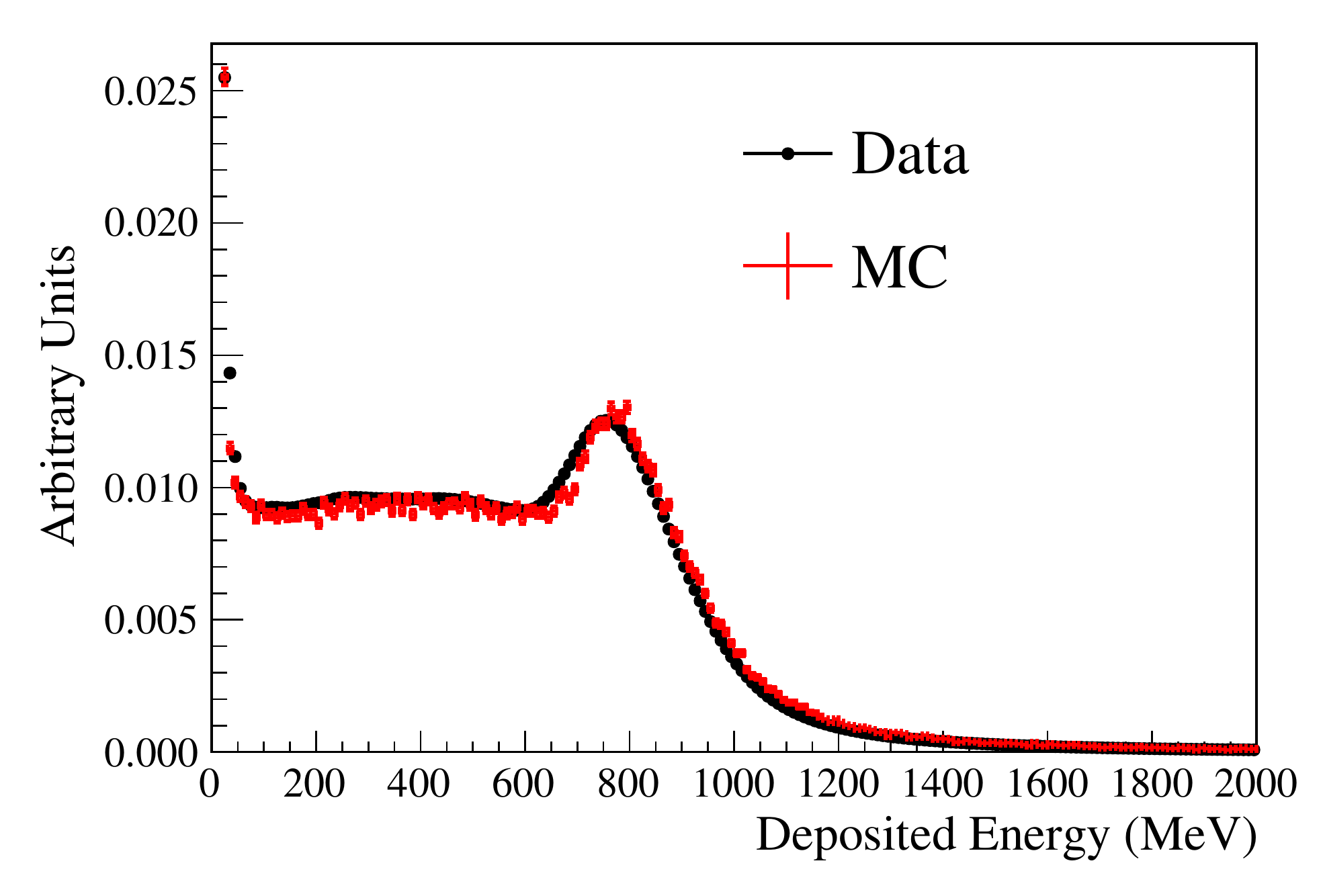}
    \caption{\label{fig:MuEdep} Distribution of energy deposited in an AD by AD-tagged muons that fall within a  [-2$\mu$s,~2$\mu$s] time window of a water-pool tagged muon in data and MC (EH1).  For data, the deposited energy is reconstructed from PMT hits.  For MC, the simulated deposited energy is shown.}
\end{figure}

In the water pool, muons are tagged by the PMT multiplicity, the number of PMTs with a signal above a threshold of 0.25 photoelectrons.  The criterion for a water-pool tagged muon is more than 12 PMTs triggered in the inner or outer water pool.  Muons passing through an AD are tagged by the amount of energy deposited in the AD. The criterion for an AD-tagged muon is an AD trigger with visible energy of at least 20~MeV.  For this analysis, AD-tagged muons that fall within a  [-2~$\mu$s,~2~$\mu$s] time window of a water-pool tagged muon are selected.  (The negative time difference is allowed to account for time offsets between detectors.)  Figure~\ref{fig:MuEdep} shows the energy deposited in an AD for selected muon events.  The peak  around 800~MeV is due to muons with path lengths of 3-4~m, corresponding to the dimensions of the GdLS region, while the peak at low energy is dominated by muons losing energy as Cherenkov light in the mineral oil.

With the criteria specified above, a sample of $N_{\mu,{\rm Obs}}$ muons is observed.
Because this sample includes muons which have traversed the AD without entering the GdLS region, a purity correction $P_{\mu}$ is applied to obtain $N_\mu$, 

\begin{equation}
\label{eqn:nmu}
 N_{\mu} = N_{\mu, {\rm Obs}}P_{\mu},
\end{equation}
where $N_{\mu}$ is the number of muons passing through the GdLS region used in Eq.~\ref{eqn:ny}. The purity $P_{\mu}$ is obtained from simulations as the ratio of the number of muons with non-zero path length in the GdLS that deposit at least 20~MeV in an AD to the total number of muons that deposit at least 20~MeV in an AD.  Table~\ref{tab:muparam} shows the values of $N_{\mu, {\rm Obs}}$ and $P_{\mu}$ for each EH.  $P_{\mu}$ is approximately 62\%; the remaining 38\% of the muons deposit at least 20~MeV by passing through the LS region only.  Muons that reach the GdLS deposit a minimum of approximately 80~MeV of energy in the LS.

\begin{figure}[htp]
\centering
    \includegraphics[width=0.49\textwidth]{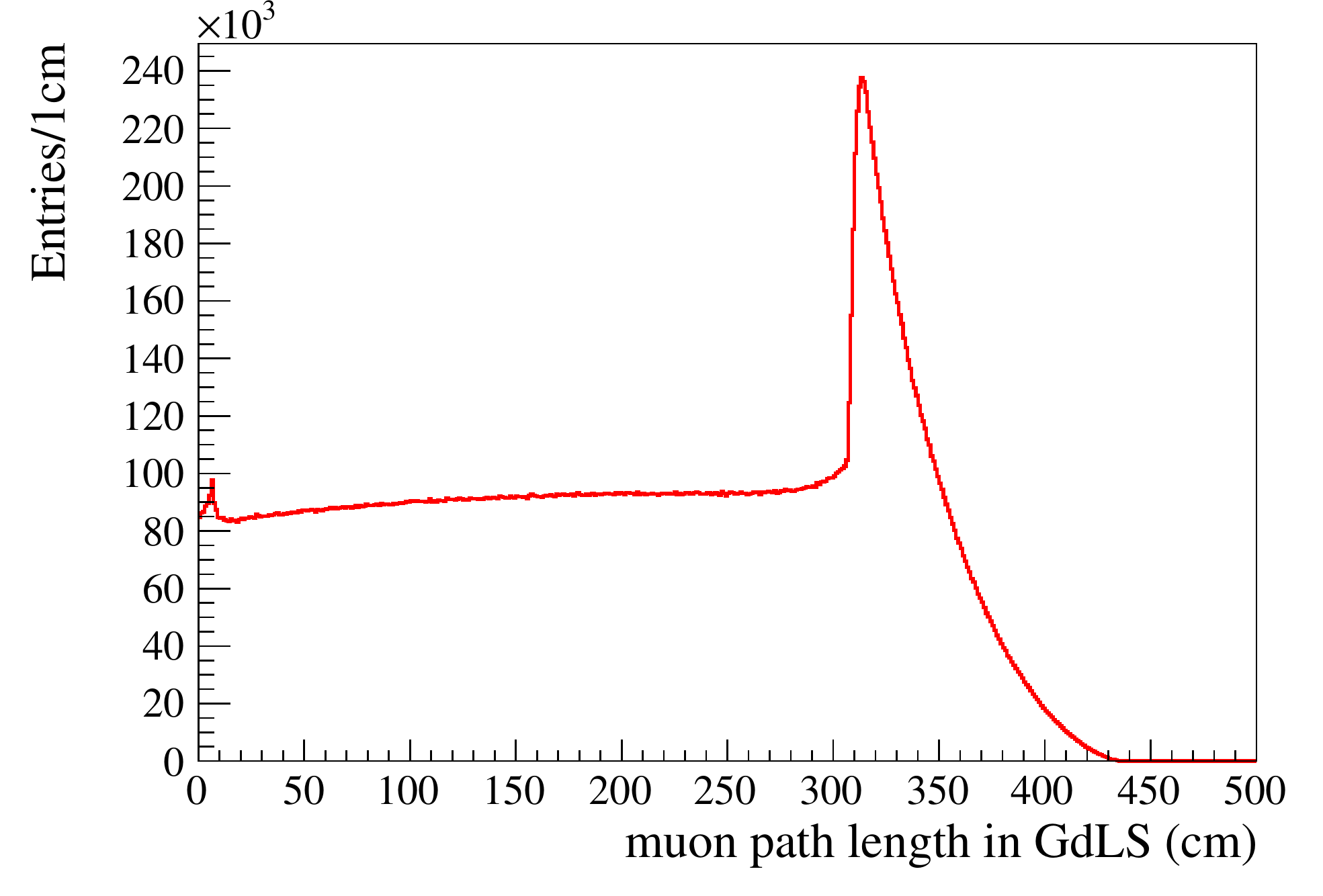}
    \caption{\label{fig:mulen} Distribution of muon path length through the GdLS from simulation (EH1).  The small peak at $\sim$6.5~cm is due to the geometry of the calibration tubes in the AD.  The large peak around 300~cm corresponds to the dimensions of the of the GdLS region.}
\end{figure}

Both $P_{\mu}$ and the average muon path length in GdLS, $L_{\rm avg}$, are geometry-related parameters that depend largely on the muon angular distribution, and therefore the values are obtained from the muon simulation.  The muon flux, energy, and angular distributions from the {\sc Music} simulation of the Daya Bay site described in Section~\ref{sec:muonsim} are input to the detector simulation.  Figure~\ref{fig:mulen} shows the distribution of muon path length through the GdLS from simulation.  The average of this distribution is used as $L_{\rm avg}$.

\begin{figure}[htp]
\centering
    \includegraphics[width=0.49\textwidth]{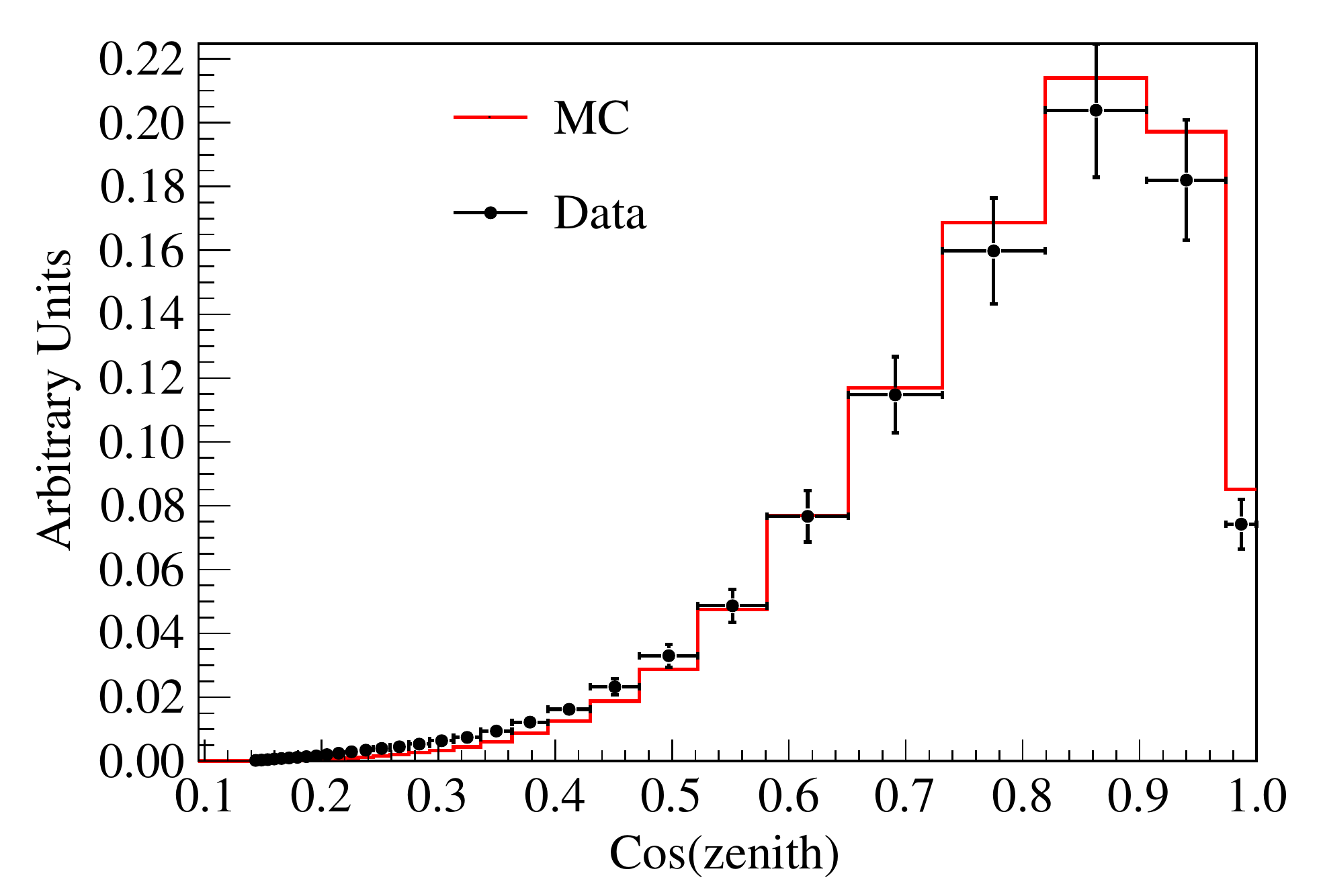}
    \includegraphics[width=0.49\textwidth]{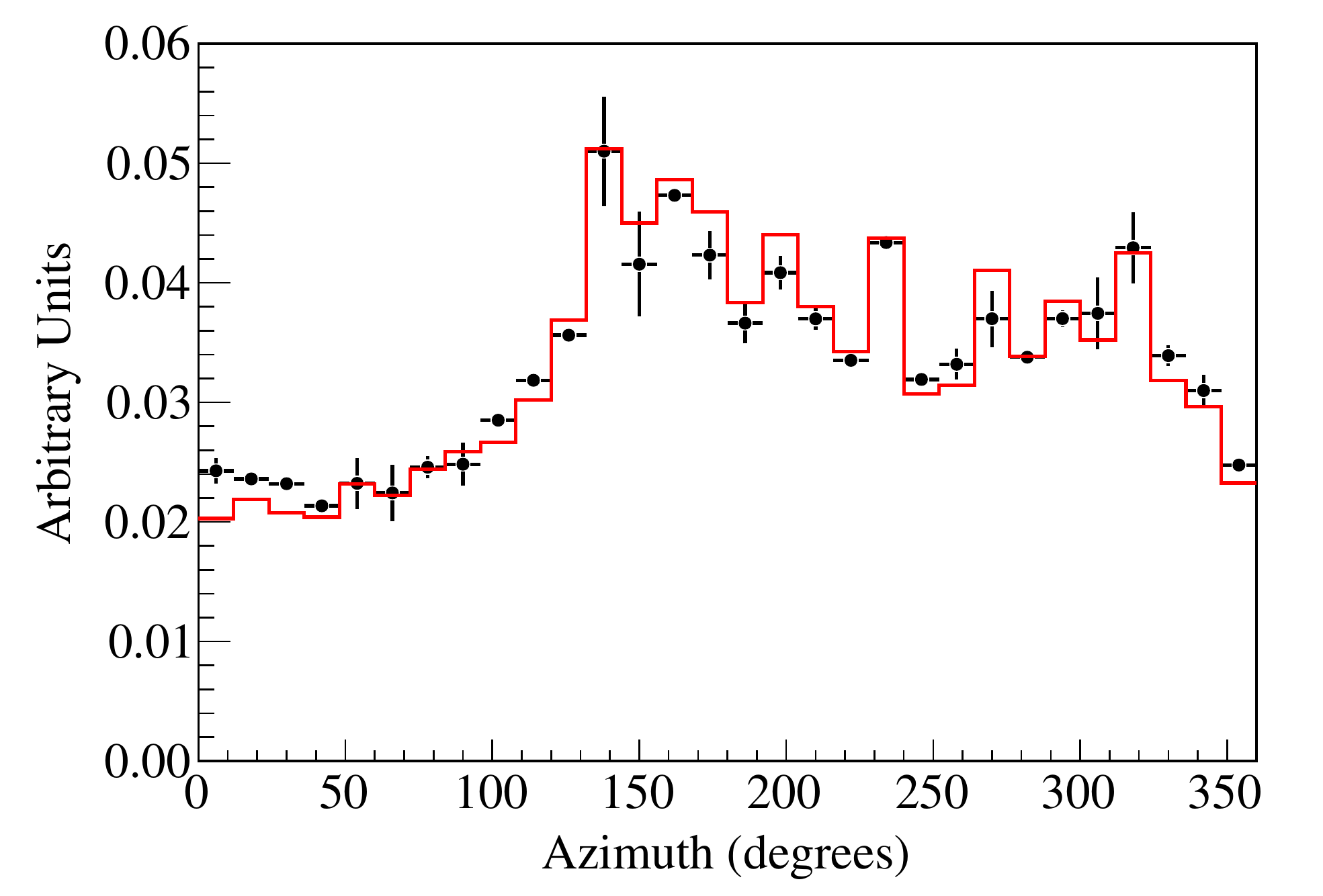}
    \caption{\label{fig:thetaRPC} Zenith angle ($\theta$) distribution (top) and azimuthal ($\phi$) distribution (bottom) from the nominal muon simulation (red line) and data (black points) for RTC events in EH1.  The corresponding distributions for EH2 and EH3 show similar agreement between the data and the nominal muon simulation.}
\end{figure}

To verify the simulated muon angular distributions, the simulated muons are compared to a sample of muons with reconstructed tracks from data.  By searching for coincident hits in the RPCs and telescope RPCs (shown in Fig.~\ref{fig:TeleRPC}), a sample of muon tracks is obtained for which the muon direction can be precisely reconstructed. Using this method, the zenith angle ($\theta$) and azimuthal angle ($\phi$) distributions of muons for these RPC Telescope Coincidence (RTC) events are obtained.  
%Muon tracks are reconstructed for RTC events as the straight line connecting the reconstructed position of the RPC hit with the reconstructed position of the RPC telescope hit.  
Figure~\ref{fig:thetaRPC} compares the angular distributions for RTC events in data and simulation.  The RTC sample is approximately 1-2\% of the total muon sample.

Because of the small angular acceptance of the telescope RPCs, the tracks in the RTC sample are a biased selection of muon tracks.  However, this sample can be used to correct the simulated total underground muon distribution based on the ratio of the measured and simulated muon distributions in the RTC sample.  The correction is done in bins of $\theta$ and $\phi$.  This data-driven or tuned prediction for the total underground muon distribution is used as an input to the detector simulation, and the tuned muon simulation is used to estimate uncertainties in $P_{\mu}$ and $L_{\rm avg}$.  The angular distributions of the RTC events for the tuned muon simulation are nearly identical to the data distributions shown in Fig.~\ref{fig:thetaRPC}.

% However, the difference in shape between data and simulation can be corrected by comparing the RTC sample from simulation to the observed RTC sample.   From the muon simulation, the ratio of the total underground muon rate and the rate for muons in the RTC sample is obtained in bins of $\theta$ and $\phi$.  Multiplying this ratio by the measured ($\theta$,$\phi$) distribution from the RTC sample in each bin gives a prediction for the total underground muon ($\theta$,$\phi$) distribution.  
   
The muon-related parameters and the associated uncertainties are summarized in Table~\ref{tab:muparam}.  The uncertainty in the product $P_{\mu}\times L_{\rm avg}$ is evaluated to account for correlations between the parameters.
%  Uncertainties are assigned to by comparing the values from the nominal muon simulation and the tuned muon simulation %(``tuned-nominal" in Table~\ref{tab:muparam}). 
The values from the nominal muon simulation are used as the central parameter values.  The difference between the nominal values and the values obtained with the tuned muon simulation is included as an uncertainty (``tuned-nominal" in Table~\ref{tab:muparam}).  The maximum difference between the values calculated using {\sc Geant4} and {\sc Fluka} is also included as an uncertainty (``{\sc Geant4}-{\sc Fluka}").   Uncertainty in the measured $\theta$ distribution, estimated conservatively at 10\%, and the MC $\phi$ distribution, estimated conservatively at 5\%, also introduce uncertainty in the parameter values from the tuned muon simulation (``$\theta$ uncertainty", ``$\phi$ uncertainty").  An additional uncertainty in $P_{\mu}$ is assigned to account for the inclusion of particles other than muons that are incorrectly tagged as muons by a 20~MeV energy deposit in the AD (``non-$\mu$").  For example, neutrons produced from showering muons in the rock surrounding the hall or water may also deposit more than 20~MeV in the AD.  Optical processes are turned off in the MC simulation, and therefore contributions to the deposited energy of the muon from Cherenkov light in the mineral oil (MO) region are not included in the simulation.  Because the deposited energy is used in the calculation of $P_{\mu}$, a systematic uncertainty is assigned to $P_{\mu}$ due to this effect (``MO dep-$E$'').  A toy MC simulation is used to calculate $P_{\mu}$ when the energy deposited in the mineral oil is included, and the difference from nominal value is used as the uncertainty.  The muon deposited energy distribution in Fig.~\ref{fig:MuEdep} has been corrected for this effect.  The statistical uncertainty is also included.
% for $L_{\textnormal{avg}}$, but is negligible for $P_{\mu}$.     
The total uncertainty is calculated as the square root of the sum of the squares of the individual uncertainties.

\begin{table}[htp]
  \centering
  \caption{\label{tab:muparam} $N_{\mu,{\rm Obs}}$ from data and $P_{\mu}$ and $L_{\textnormal{avg}}$  from the nominal muon simulation for all three experimental halls. 
%The uncertainties in each value ($\delta P_{\mu}$ and $\delta L_{\rm avg}$) due to various effects are also shown, as well as 
The relative uncertainties in the combined parameter $P_{\mu}\times L_{\rm avg}$ due to different sources are shown.}
  \begin{tabular}{lccc}
    \hline
    \hline
 & EH1 & EH2 & EH3\\
    \hline 
$N_{\mu,{\rm Obs}}$ & $2.07 \times 10^{9}$ & $1.29\times 10^{9}$ &$1.87\times 10^{8}$ \\
\hline
$P_{\mu}$ nominal & 62.36\% & 62.40\% & 62.42\% \\
%$\delta P_{\mu}$ ({\sc Geant4}-{\sc Fluka}) & 2.78\% & 2.80\% & 2.02\%\\
%$\delta P_{\mu}$ (tuned-nominal) & 0.70\% & 0.50\% & 0.30\%\\
%$\delta P_{\mu}$ ($\theta$ uncertainty) & 0.30\% & 0.30\% & 0.30\%\\
%$\delta P_{\mu}$ ($\phi$ uncertainty) & 0.66\% & 0.66\% & 0.66\%\\
%$\delta P_{\mu}$ (non-$\mu$) & 0.64\% & 0.50\% & 0.75\%\\
%$\delta P_{\mu}$ (MO dep-$E$) & 1.45\% & 1.45\% & 1.45\%\\
%$\delta P_{\mu}$ (total) & 3.34\% & 3.31\% & 2.71\%\\
%\hline
$L_{\textnormal{avg}}$ nominal &204.1~cm & 204.5~cm & 204.9~cm\\
%$\delta L_{\textnormal{avg}}$ ({\sc Geant4}-{\sc Fluka}) & 0.6~cm & 0.6~cm & 1.3~cm \\
%$\delta L_{\textnormal{avg}}$ (tuned-nominal) & 0.3~cm & 0.9~cm & 0.4~cm \\
%$\delta L_{\textnormal{avg}}$ ($\theta$ uncertainty) & 1.0~cm & 1.0~cm & 1.0~cm \\
%$\delta L_{\textnormal{avg}}$ ($\phi$ uncertainty) & 0.5~cm & 0.5~cm & 0.5~cm \\
%$\delta L_{\textnormal{avg}}$ (statistical) & 0.2~cm &0.8~cm & 0.3~cm\\
%$\delta L_{\textnormal{avg}}$ (total) & 1.3~cm &1.8~cm & 1.8~cm\\
%\hline
%$P_{\mu} \times L_{ \rm avg}$ nominal & 127.3~cm & 127.6~cm & 127.9~cm\\
$\delta (P_{\mu}  L_{\rm avg})$ ({\sc Geant4}-{\sc Fluka}) & 4.71\% & 4.78\% & 2.66\%\\
$\delta (P_{\mu}  L_{\rm avg})$ (tuned-nominal) & 1.18\% & 1.25\% & 0.23\%\\
$\delta (P_{\mu}  L_{\rm avg})$ ($\theta$ uncertainty) & 0.08\% & 0.08\% & 0.08\%\\
$\delta (P_{\mu}  L_{\rm avg})$ ($\phi$ uncertainty) & 0.71\% & 0.71\% & 0.70\%\\
$\delta (P_{\mu}  L_{\rm avg})$ (non-$\mu$) & 1.03\%  & 0.80\%  & 1.20\%\\
$\delta (P_{\mu}  L_{\rm avg})$ (MO dep-$E$) & 2.33\% & 2.32\% &2.32\% \\
$\delta (P_{\mu}  L_{\rm avg})$ (statistical) & 0.10\% & 0.39\% & 0.15\%\\
$\delta (P_{\mu}  L_{\rm avg})$ (total) & 5.53\% & 5.58\% & 3.80\%\\
\hline
\hline
 \end{tabular}
  \centering
\end{table}

\subsection{Neutron event selection}
\label{sect:nsel}

To determine the number of neutrons produced due to muons passing through the GdLS, neutron captures on Gd following a muon signal are selected.  Simulations show that the average kinetic energy of a neutron produced in the GdLS by a muon is around 40~MeV, with a long tail that extends to around 1~GeV.  Neutrons travel an average of $\sim$40~cm before capturing on Gd.  
%Approximately one-quarter of the neutrons escape the GdLS without being captured. 

\begin{figure}[htp]
\centering
    \includegraphics[width=0.49\textwidth]{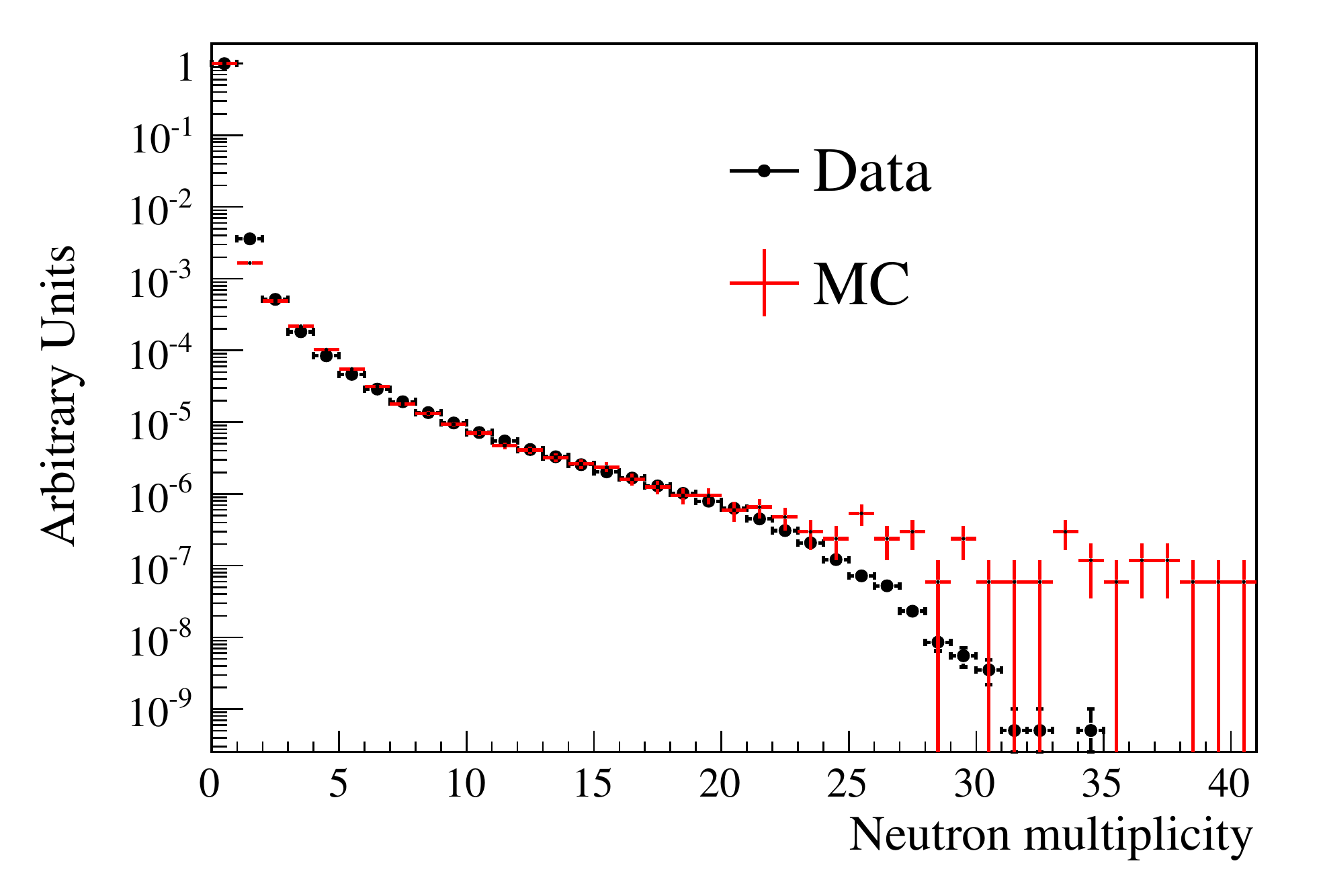}
    \caption{\label{fig:nmult} Comparison of neutron multiplicity in data and MC in EH1.  For the MC, true neutron captures on Gd are selected between 10 and 200~$\mu$s after an AD-tagged muon.  For the data, neutron captures are selected with a 6-12~MeV energy range between 10 and 200~$\mu$s after an AD-tagged muon.  To suppress random background in the data, no other AD-tagged muon is allowed in a [-0.5,0.5]~ms window.  This distribution has been corrected for readout window efficiency and the effect of blocked triggers.}
\end{figure}

To select the Gd captures, AD triggers are chosen with energy between 6 and 12~MeV occurring in a signal time window, at least 10~$\mu$s and no more than 200~$\mu$s, after an AD-tagged muon.  Triggers occurring $<$10~$\mu$s after a muon are not used due to afterpulsing and ringing in the PMTs following the passage of a muon.  This criterion also vetoes decay electrons from stopping muons.  Figure~\ref{fig:nmult} compares the neutron multiplicity between data and MC, where the neutron multiplicity is defined as the number of AD triggers after an AD-tagged muon that satisfy the Gd capture criteria.  This distribution has been corrected for readout window efficiency and the effect of blocked triggers, discussed later in this section.  Studies indicate that cases where multiple muons pass through the detector before nGd capture candidates are rare and can be neglected in this analysis.

\begin{figure}[htp]
\centering
    \includegraphics[width=0.49\textwidth]{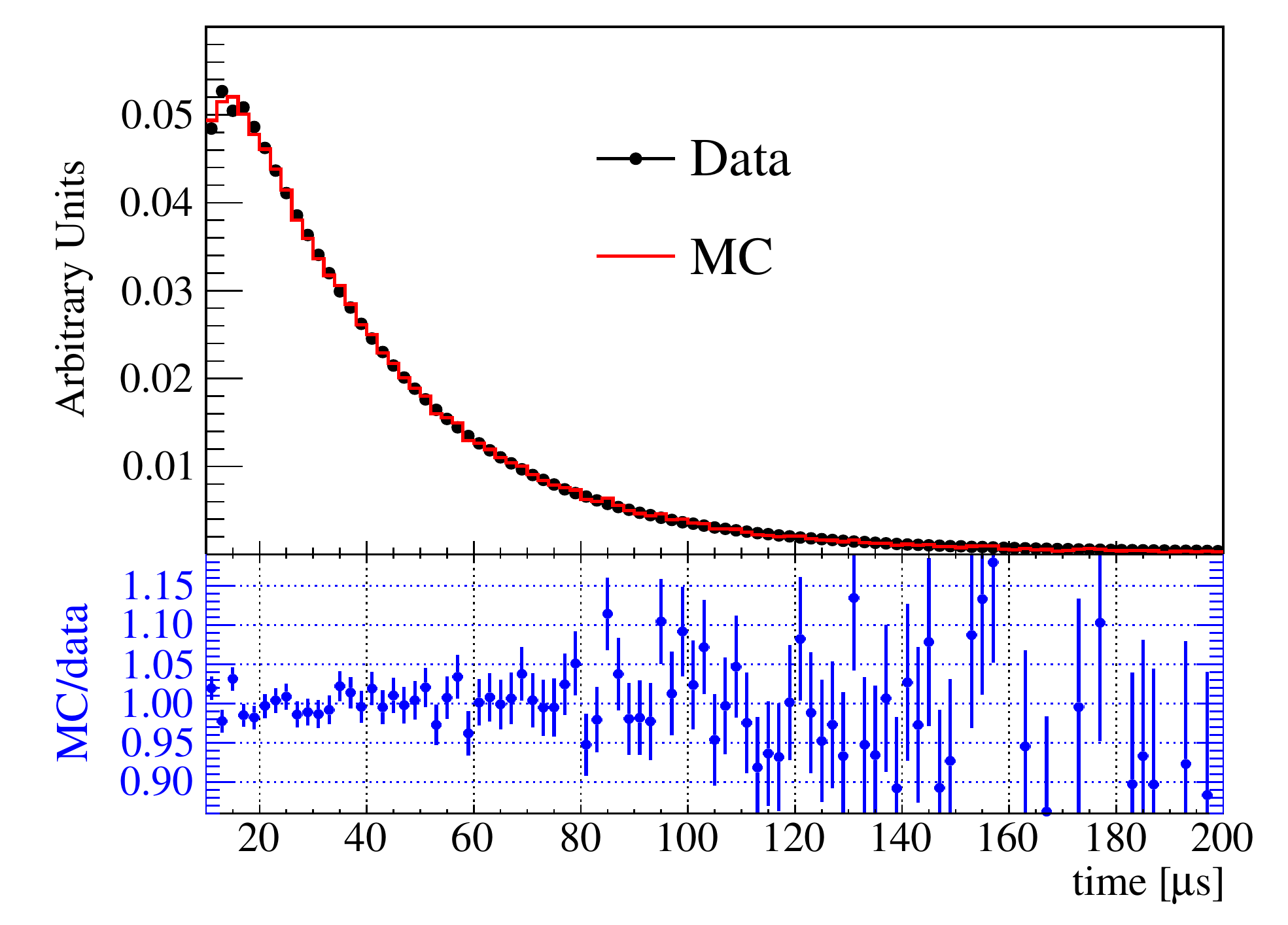}
    \caption{\label{fig:neutronCaptime} Time between the muon and delayed events (neutron capture time) for data and MC in EH1.}
\end{figure}

\begin{figure}[htp]
\centering
    \includegraphics[width=0.49\textwidth]{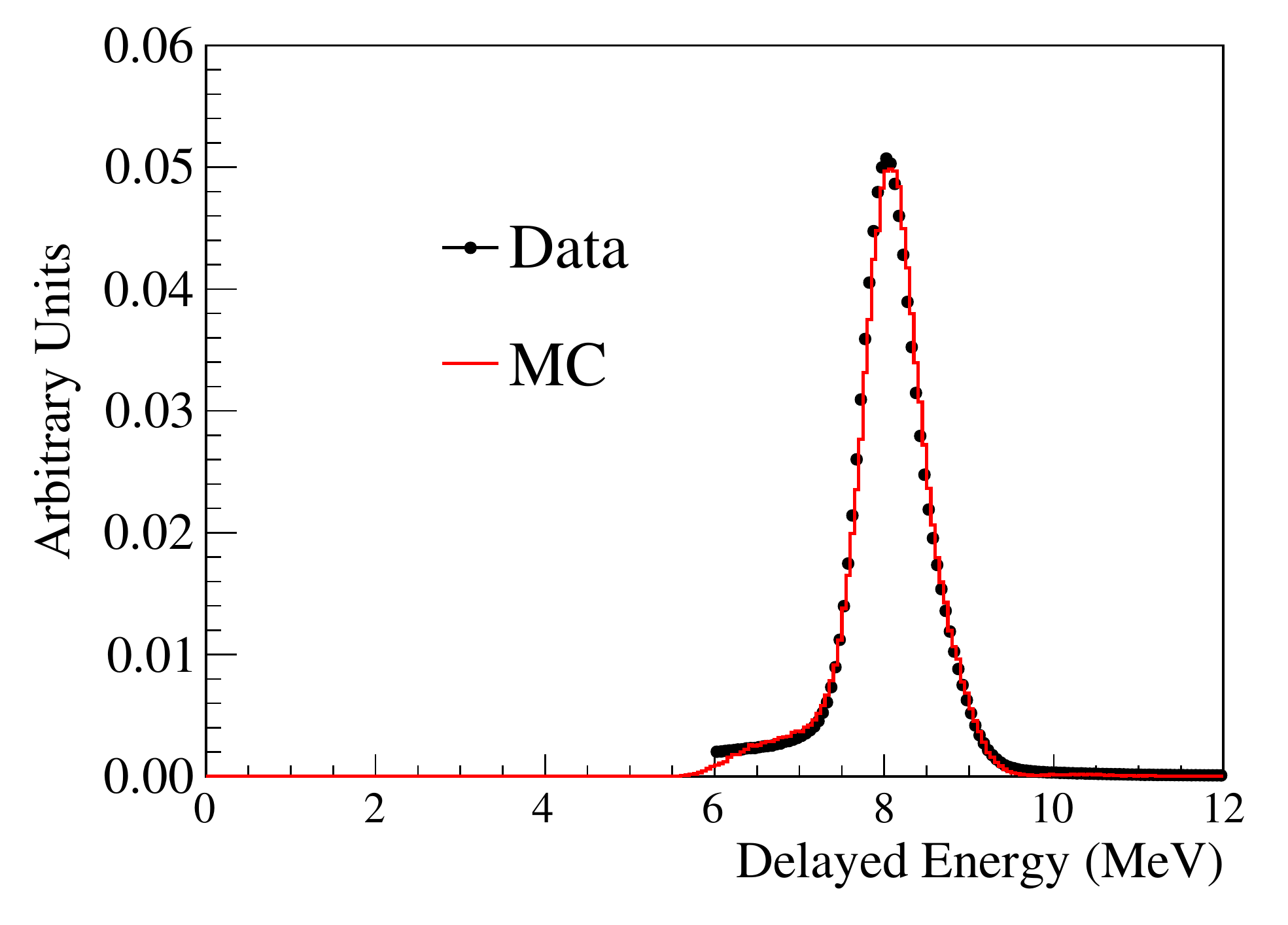}
    \caption{\label{fig:neutronSpectrum} Energy of the delayed events (neutron capture energy) for data and MC in EH1.}
\end{figure}

The selected events in the signal time window  include random backgrounds unrelated to the muon passage.  These backgrounds are estimated by selecting AD triggers with energy between 6 and 12~MeV occurring long after the muon passage, in a time window between 1010~$\mu$s and 1200~$\mu$s after the muon.  Because the neutron capture time is $\sim$30~$\mu$s, the contribution of neutron captures in this late time window is negligibly small. Given that both cosmic muons and background events are distributed randomly in time, Poisson statistics dictates that the distribution of background events measured in time since the last muon is an exponential with a time dependence on the muon rate, $R_{\mu}$.
%Based on Poisson distributions for cosmic muons and background events that are distributed randomly in time, the distribution of background events measured in time since the last muon is an exponential with a time dependence on the muon rate, $R_{\mu}$. 
Therefore, the selected background events in the late time window slightly underestimates the background contribution to the selected events in the signal time window. A parameter $\alpha$ is used to apply a small correction to the number of events in the background window to take this into account.  The number of selected neutron captures ($N_{\rm cap}$) is given by

\begin{equation}
\label{eqn:ncapsel}
 N_{\textnormal{cap}} =N_{10-200~\mu\rm{s}}-\alpha N_{1010-1200~\mu\rm{s}},
\end{equation}
where $N_{10-200~\mu\rm{s}}$ is the number of selected events in the signal time window, $N_{1010-1200~\mu\rm{s}}$ is the number of selected events in the background time window, and $\alpha$ is the correction factor.  The correction factor $\alpha$ is given by

\begin{equation}
\alpha = \frac{\int_{10~\mu\rm{s}}^{200~\mu\rm{s}} e^{-R_{\mu}t}dt}{\int_{1010~\mu\rm{s}}^{1200~\mu\rm{s}} e^{-R_{\mu}t}dt},
\end{equation}
where $R_{\mu}$  is the measured rate of AD muons (1.21$\pm$0.12~Hz, 0.87$\pm$0.09~Hz, and 0.056$\pm$0.006~Hz in EH1, EH2, and EH3, respectively~\cite{Dayabay:2014vka}).
The number of selected events in the signal and background time windows and the value of $\alpha$ are shown in Table~\ref{tab:nsel}.  Figures~\ref{fig:neutronCaptime} and \ref{fig:neutronSpectrum} compare data and MC for distributions of candidate neutron captures.  Figure~\ref{fig:neutronCaptime} shows the distribution of time between the muon and delayed events for both data and MC.  Figure~\ref{fig:neutronSpectrum} shows the energy of the delayed events.  The background subtraction has been applied in both figures.

\begin{table}[htp]
  \centering
  \caption{\label{tab:nsel} Selected events (in millions) in the signal and background time windows and the background correction factor $\alpha$ for each experimental hall.}
  \begin{tabular}{lccc}
    \hline
    \hline
     & EH1 & EH2 & EH3 \\
\hline
$N_{10-200~\mu\rm{s}}$ & ~$14.2$ & $8.84$ & $2.00$ \\
$N_{1010-1200~\mu\rm{s}}$ & ~$0.367$ & $0.169$ & $0.00259$\\
\hline
$\alpha$ & 1.02 & 1.01  & 1.00 \\
    \hline
    \hline
  \end{tabular}
  \centering
\end{table}

A systematic uncertainty in the number of selected neutron captures ($N_{\rm cap}$) is assigned due to blocked triggers.  When the event rate is high, the electronics buffer can become saturated, and any trigger that occurs during this time will be blocked.  
%When the buffer clears, the number of blocked triggers is recorded in the next trigger.
 Blocked triggers can occur when a muon suffers a large energy loss ($>4$~GeV) in the AD, which generates many triggers, including neutrons.  However, there is no way to determine if the blocked triggers are neutron captures. To be conservative, all blocked triggers are assumed to be neutron captures, and the systematic uncertainty is calculated from data as the number of selected neutrons that have blocked triggers in the time window since last muon relative to the total number of selected neutrons. Table~\ref{tab:blocked} gives the uncertainty in $N_{cap}$ due to this effect in each experimental hall.

\begin{table}[htp]
  \centering
  \caption{\label{tab:blocked} Relative uncertainty in $N_{cap}$ due to blocked triggers for each experimental hall.}
  \begin{tabular}{lccc}
    \hline
    \hline
     & EH1 & EH2 & EH3 \\
\hline
$\delta N_{\rm cap}/N_{\rm cap}$ & ~$0.50$\% & $0.50$\% & $1.3$\% \\
\hline
    \hline
  \end{tabular}
  \centering
\end{table}

The number of selected neutron captures from Eq.~\ref{eqn:ncapsel} is further corrected for the efficiency of the selection criteria, including the energy selection efficiency $\varepsilon_E$, the time window selection efficiency $\varepsilon_t$, the fraction of neutrons captured on Gd $\varepsilon_{\rm Gd}$, and the electronics readout window efficiency $\varepsilon_{\rm ro}$.  The values and uncertainties for $\varepsilon_E$ and $\varepsilon_{\rm Gd}$ are taken from other Daya Bay analyses~\cite{An:2015nua,An:2016srz} and are the same for every AD.  Because the time window to select neutron captures is different in this analysis, the value of $\varepsilon_t$ is calculated from the simulation of cosmogenically-produced neutrons. The uncertainty in the value of $\varepsilon_t$ is estimated by comparing the IBD neutron capture time distribution in data and MC.  The electronics readout window efficiency $\varepsilon_{\rm ro}$ corrects for the fact that within the 1.2-$\mu$s electronics readout window, only the first neutron capture will be read out by the electronics.  For high multiplicity events, any subsequent neutron captures within that 1.2-$\mu$s readout time will be lost and not counted in the analysis.  (The neutron multiplicity distribution in Fig.~\ref{fig:nmult} has been corrected for this effect and the effect of blocked triggers.)  The readout window efficiency factor is calculated from the simulation as the ratio of the number of neutron captures that would be selected (because their captures occur first in the time window) to the total number of neutron captures regardless of their timing.  The uncertainty in the efficiency is calculated by comparing values of the neutron yield calculated using different-sized time windows to select the signal and background neutrons following a muon.  The maximum fractional difference in the yield from nominal is taken as the relative uncertainty in the readout window efficiency. The values and uncertainties for all of the efficiency corrections are summarized in Tables~\ref{tab:cuteffi} and \ref{tab:cuteff2}.

\begin{table}[htp]
\caption{\label{tab:cuteffi} Efficiency of neutron capture selection due to the energy cut, time cut, and Gd-capture fraction (same for each EH)}
\begin{tabular}{lcc}
\hline
\hline
& Efficiency ($\varepsilon$) & Uncertainty ($\delta_{\varepsilon}/\varepsilon$)\\
\hline
Gd-capture fraction ($\varepsilon_{\rm Gd}$) & 85.4\% & 0.4\%\\
Energy ($\varepsilon_E$) & 92.71\% & 0.97\%\\
Time since muon ($\varepsilon_t$) & 83.7\% & 0.3\%\\
\hline
\hline
\end{tabular}
\end{table}

\begin{table}[htp]
\caption{\label{tab:cuteff2} Efficiency of neutron capture selection due to electronics readout, $\varepsilon_{\rm ro}$, calculated for each EH}
\begin{tabular}{lcc}
\hline
\hline
& Efficiency ($\varepsilon_{\rm ro}$) & Uncertainty ($\delta_{\varepsilon_{\rm ro}}/\varepsilon_{\rm ro}$)\\
\hline
EH1 & 89.7\% & 2.21\%\\
EH2 & 89.7\% & 2.21\%\\
EH3 & 86.8\% & 1.56\%\\
\hline
\hline
\end{tabular}
\end{table}

Neutrons that are captured in the stainless steel vessel (SSV) instead of Gd are included in the sample if the emitted gammas enter the LS or GdLS and produce a signal that satisfies the selection cuts.    A correction is applied to account for the inclusion of these neutron captures in the sample.  The SSV correction, $f_{\textnormal{nSSV}}$, is the ratio of the number of neutrons captured in the SSV to the total number of neutron captures, obtained from simulation.  The values of this correction factor are shown in Table~\ref{tab:npurity}.

Another source of contamination is neutrons selected in time with a signal that is identified as a muon, but is actually some other type of particle.
% an incorrect muon tag, which is a 20~MeV or more energy deposit in an AD created by something other than a muon. 
%For example,
Secondary particles from a showering muon in the rock or water could deposit 20~MeV or more in the AD, causing the event to be incorrectly tagged as a muon, and the subsequent neutron captures are incorrectly included in the sample.   To account for this effect, a correction $f_{\textnormal{non-}\mu}$ is applied, calculated as the ratio from simulation of the number of neutrons captured on Gd following a``non-$\mu$" ($\geq$~20~MeV energy deposit, but not a muon) to the number of neutrons captured on Gd following any event tagged as a muon (any $\geq$~20~MeV energy deposit).  
%The value of $f_{\textnormal{non-}\mu}$ is obtained from simulation.

Table~\ref{tab:npurity} summarizes the correction values and their uncertainties for each EH.  The maximum difference between the values calculated using {\sc Geant4} and {\sc Fluka} (``{\sc Geant4}-{\sc Fluka}") is used for the uncertainty in $f_{\textnormal{nSSV}}$.
The uncertainty in $f_{\textnormal{non-}\mu}$ is determined using the difference between the values calculated using {\sc Geant4} and {\sc Fluka}, plus a smaller uncertainty due to neutron propagation (``n propagation") in the MC, which will be described below.  Small statistical uncertainties in both parameters are also included.  The total uncertainty for each parameter is the square root of the sum of the squares of the individual uncertainties.

\begin{table}[htp]
  \centering
  \caption{\label{tab:npurity} Neutron capture sample correction factors for each experimental hall. The uncertainties in each value ($\delta f_{\rm nSSV}$ and $\delta f_{{\rm non}-\mu}$) due to various effects are also shown.}
  \begin{tabular}{lccc}
    \hline
    \hline
     & EH1 & EH2 & EH3 \\
\hline
 $f_{\textnormal{nSSV}}$ nominal  & 3.58\% & 3.96\% & 3.74\% \\
$\delta f_{\textnormal{nSSV}}$ ({\sc Geant4}-{\sc Fluka}) & 0.64\% & 0.64\% & 0.64\%\\
$\delta f_{\textnormal{nSSV}}$ (statistical) & 0.05\% & 0.15\% & 0.05\%\\
$\delta f_{\textnormal{nSSV}}$ (total) & 0.64\% & 0.66\% & 0.64\%\\
\hline
 $f_{\textnormal{non-}\mu}$ nominal & 5.17\% & 4.94\% & 5.35\%\\
 $\delta f_{\textnormal{non-}\mu}$ ({\sc Geant4}-{\sc Fluka}) & 2.24\% & 2.34\% & 1.95\% \\
 $\delta f_{\textnormal{non-}\mu}$ (n propagation) & 0.92\% & 0.88\% & 0.95\% \\
 $\delta f_{\textnormal{non-}\mu}$ (statistical) & 0.05\% & 0.15\% & 0.05\%\\
 $\delta f_{\textnormal{non-}\mu}$ (total) & 2.42\% & 2.50\% & 2.17\%\\
    \hline
    \hline
  \end{tabular}
  \centering
\end{table}

With the efficiency and purity corrections described above, the corrected number of neutron captures on Gd following a GdLS muon is 
\begin{align}
\label{eqn:ngd}
N_{\textnormal{nGd}} =N_{\textnormal{cap}} \times \frac{(1-f_{\textnormal{nSSV}})~(1-f_{\textnormal{non}-\mu})}{\varepsilon_{\textnormal{Gd}}~\varepsilon_E~\varepsilon_t~\varepsilon_{\rm ro}}
\end{align}

The final step is to determine the number of neutrons produced due to muons passing through the GdLS based on the number of neutron captures on Gd given by Eqn.~\ref{eqn:ngd}.  Two parameters, $R_{\textnormal{spill}}$ and $R_{\textnormal{det}}$, are used to determine this relationship. 
$R_{\textnormal{spill}}$ accounts for the net effect of spill-in, where neutrons produced by muons outside the GdLS are detected via Gd capture, and spill-out, where neutrons produced by muons in the GdLS escape before capture on Gd.  The value of $R_{\textnormal{spill}}$, obtained from simulation, is the ratio of the number of neutrons produced inside the GdLS to the number of neutrons captured on Gd.  In EH1, the spill-out correction is approximately 26\%, while the spill-in correction is around 20\%, leading to a net correction of 6\%.  The values for $R_{\rm spill}$ for each EH are shown in Table~\ref{tab:R}.

The parameter $R_{\textnormal{det}}$ accounts for the finite detector size.  Some neutrons are neither produced nor captured inside the GdLS, but are indirectly produced by the passage of a muon through the GdLS.  For example, a muon passing through the GdLS emits a gamma, which leaves the detector and subsequently produces a neutron that is not detected in the AD.  An arbitrarily large GdLS detector would tag this neutron and associate it to the muon, but it goes undetected in this analysis due to the AD size. Therefore, this correction is necessary for consistency with the definition of neutron yield used in other experiments with different detector geometries. The value of $R_{\textnormal{det}}$, obtained from simulation, is the ratio of the number of neutrons produced inside the GdLS to the number of neutrons produced due to a muon's passage through the GdLS (regardless of the generation point of the neutron).

With these corrections, the number of neutrons produced due to the passage of a GdLS muon can be found with

\begin{equation}
\label{eqn:nn}
N_{n} =\frac{R_{\textnormal{spill}}}{R_{\textnormal{det}}}\times N_{\textnormal{nGd}}.
\end{equation}

\begin{figure}
 \centering
 \includegraphics[width=0.49\textwidth]{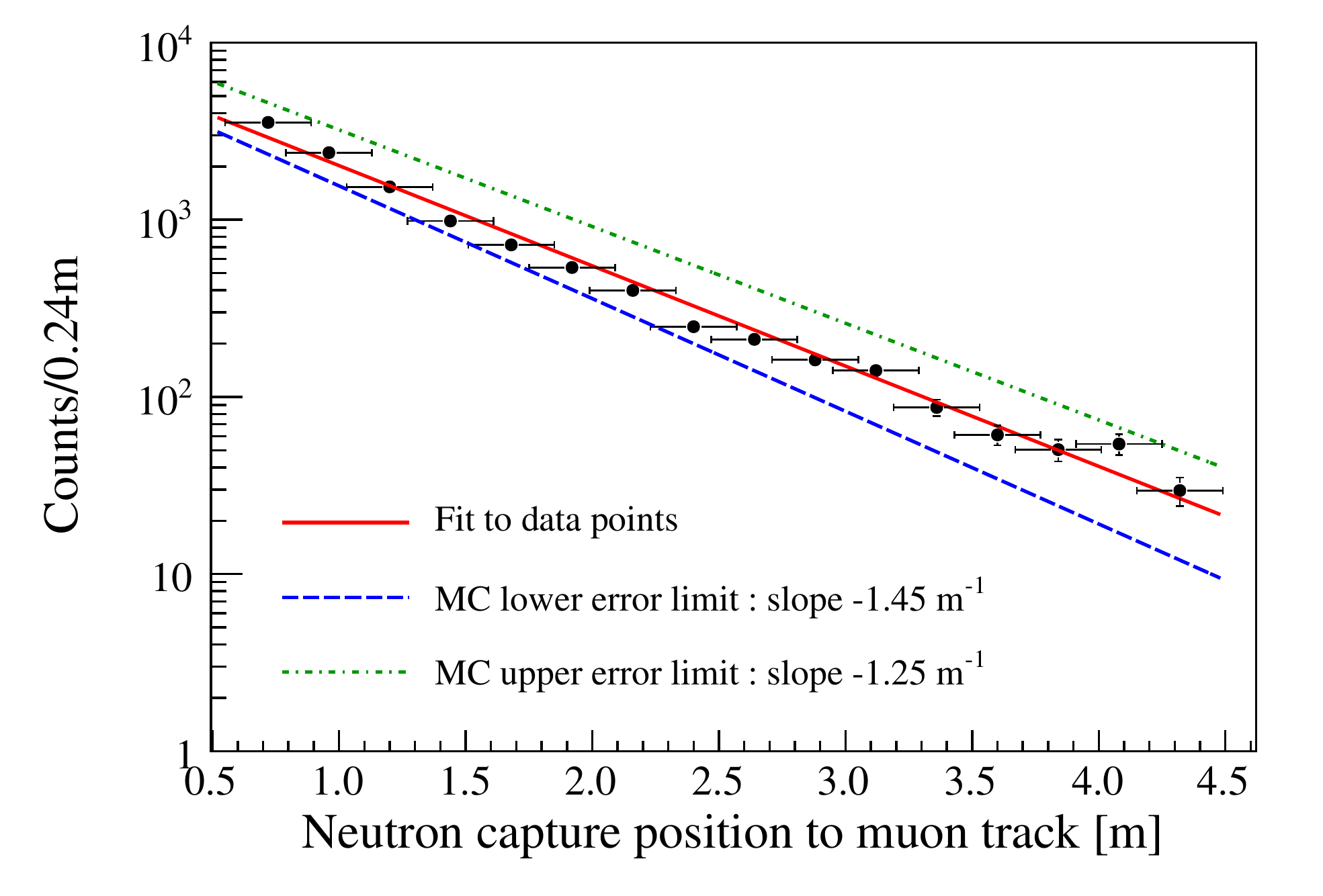}
  \caption{\label{fig:neutron_distance} Semi-log distribution of the perpendicular distance between neutron capture position and the RTC event muon track for EH1 data.  The best linear fit of the logarithm of the number of counts as a function of the distance for the data  is shown, in addition to the lines drawn with the upper and lower limit slopes used to determine the neutron energy scaling in the MC.}
\end{figure}

The values for $R_{\textnormal{spill}}$ and $R_{\textnormal{det}}$ and the associated uncertainties are summarized in Table~\ref{tab:R}.  The uncertainty in the ratio $R_{\rm spill}/R_{\rm det}$ is evaluated to account for correlations between the parameters. The values $R_{\textnormal{spill}}$, $R_{\textnormal{det}}$, and  $f_{\textnormal{non-}\mu}$ depend on the neutron propagation model in the simulation.  
Uncertainties in these parameters are estimated by comparing neutron propagation in data and simulation. Neutron captures associated with muon tracks from the RTC sample are selected.  The neutron capture position is reconstructed with an uncertainty of approximately 20~cm using the method described in Ref.~\cite{An:2016ses}.  Figure~\ref{fig:neutron_distance} shows the distribution of the minimum distance between the reconstructed neutron capture position and muon track in semi-log scale for the EH1 data.  A linear fit to the logarithm of the number of counts as a function of the distance performed on the data provides an allowed range for the slope from $-1.45$ to $-1.25$~m$^{-1}$.  The neutron propagation in the simulation is modified by applying a scaling factor to the simulated neutron energy.  In the simulation, slopes of $-1.45$ and $-1.25$~m$^{-1}$ for the same distribution are obtained when the neutron energy is scaled by factors of 0.87 and 1.83, respectively.
The parameters $R_{\rm spill}$, $R_{\rm det}$, and  $f_{\textnormal{non-}\mu}$ are calculated in the simulation with the  neutron energy scaling factor set to 0.87 and then set to 1.83.  This provides a range in the values of $R_{\textnormal{spill}}/R_{\textnormal{det}}$ and  $f_{\textnormal{non-}\mu}$ which is used as the uncertainty. The same process is applied to the other EHs.  This is the dominant source of uncertainty in the neutron yield.   The maximum difference between the values of $R_{\rm spill}$ and $R_{\rm det}$ calculated using {\sc Geant4} and {\sc Fluka} is also included as an uncertainty, as well as the statistical uncertainty.

\begin{table}[htp]
  \centering
  \caption{\label{tab:R} $R_{\rm spill}$ and $R_{det}$, factors that relate the number of neutrons produced due to a muon's trajectory through the GdLS to the number of observed neutron captures on Gd.  The relative uncertainties
% in each value ($\delta R_{\rm spill}$ and $\delta R_{\rm det}$) due to various effects are also shown, as well as the relative uncertainty
 in the combined parameter $R_{\rm spill}/ R_{\rm det}$
are shown.}
  \begin{tabular}{lccc}
    \hline
    \hline
     & EH1 & EH2 & EH3 \\
\hline
$R_{\textnormal{spill}}$~nominal & 1.062 & 1.054 & 1.065 \\
$R_{\textnormal{det}}$~nominal  & 0.970 & 0.972 & 0.964 \\
%$R_{\textnormal{spill}}/ R_{\textnormal{det}}$~nominal & 1.095 & 1.085 & 1.105 \\
$\delta (R_{\textnormal{spill}}/ R_{\textnormal{det}})$~({\sc Geant4}-{\sc Fluka}) & 2.01\% & 2.03\% & 1.99\% \\
$\delta (R_{\textnormal{spill}}/ R_{\textnormal{det}})$~(n propagation) & 4.66\% & 4.70\% & 4.62\%\\
$\delta (R_{\textnormal{spill}}/ R_{\textnormal{det}})$~(statistical) & 0.42\% & 1.26\% & 0.42\%\\
$\delta (R_{\textnormal{spill}}/ R_{\textnormal{det}})$~(total) & 5.09\% & 5.27\% & 5.05\%\\
    \hline
    \hline
  \end{tabular}
  \centering
\end{table}

%A study performed to understand the effect that the choice of signal time window has on the results found a maximum difference of approximately $0.1\times 10^{-5}$~$\mu^{-1}$g$^{-1}$cm$^2$ in the yield.  This is included in the uncertainty in the neutron yield.  The remaining uncertainty is from the uncertainties in the parameters described in the previous sections.

\section{Results}

The neutron yield calculated by Eq.~\ref{eqn:ny} at each EH is shown in Table~\ref{tab:NY}.  
$E^{\mu}_{\textnormal{avg}}$ is the average muon energy for muons passing through the GdLS at each EH, calculated from the {\sc Music} simulation.  The uncertainty in the average muon energy predicted by {\sc Music} is about 6\%, dominated by uncertainties in the mountain profile and rock density.  The neutron yields predicted from {\sc Geant4} and {\sc Fluka} simulations at each EH are also shown, with statistical uncertainties only. 

\begin{table}[htp]
\caption{\label{tab:NY} Measured and predicted neutron yield for each EH in units of $\times 10^{-5}\mu^{-1}~{\rm g}^{-1}~{\rm cm}^2$.  The measured value is determined from Eq.~\ref{eqn:ny} using the corrections described in previous sections.  The predicted values from {\sc Geant4} and {\sc Fluka} are obtained by counting neutrons produced due to simulated muons passing through the GdLS assuming a realistic muon flux and detector geometry.  The average muon energy from the {\sc Music} simulation in each EH is also given.}
\begin{tabular}{lccc}
\hline
\hline
 & EH1 & EH2 & EH3\\
\hline
 $E^{\mu}_{\textnormal{avg}}$~(GeV) & 63.9 $\pm$3.8 & 64.7 $\pm$3.9 & 143.0 $\pm$8.6\\
\hline
\multicolumn{4}{l}{Measured Values ($\times 10^{-5}\mu^{-1}~{\rm g}^{-1}~{\rm cm}^2$)}\\
 $Y_{n}$  & 10.26 $\pm$0.86 & 10.22 $\pm$0.87 & 17.03 $\pm$1.22 \\
\hline
\multicolumn{4}{l}{MC Predictions ($\times 10^{-5}\mu^{-1}~{\rm g}^{-1}~{\rm cm}^2$)}\\
$Y_{n}$~({\sc Geant4})~~& 7.53 $\pm$0.01  & 7.47 $\pm$ 0.05 & 13.35 $\pm$ 0.03 \\
$Y_{n}$~({\sc Fluka})& ~~8.34  $\pm$0.02~~  & ~~8.70  $\pm$0.03~~ & ~~17.15 $\pm$0.04~~ \\
\hline
\hline
\end{tabular}
\end{table}

\subsection{Comparison with other experiments}
\begin{figure*}[htp]
\centering
\includegraphics[width=0.75\textwidth]{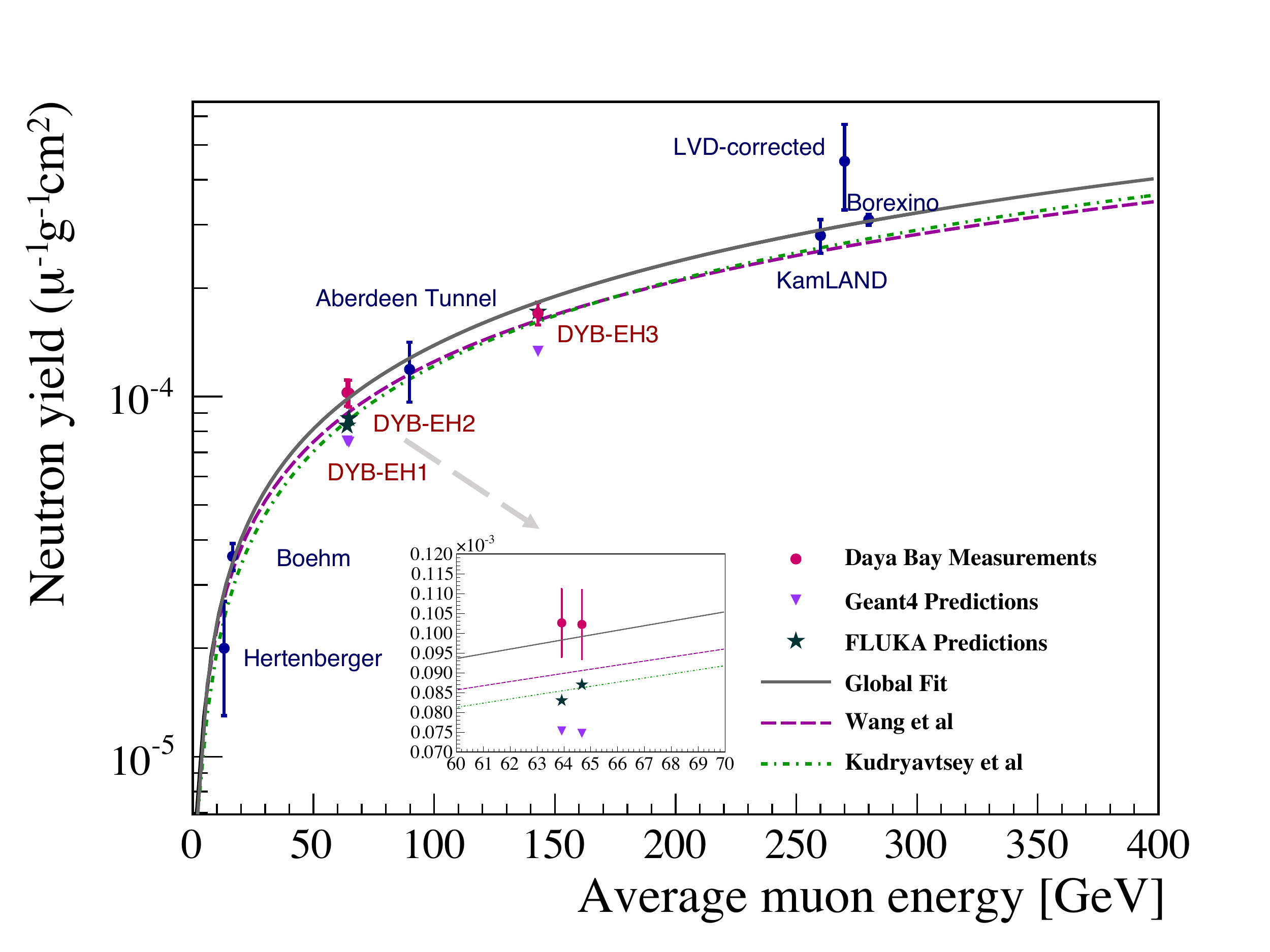}
\caption{\label{fig:Compare} Neutron yield vs. average muon energy from the three Daya Bay experimental halls compared to other experiments.  The points for Daya Bay EH1 and EH2, which differ in energy by less than 1 GeV, are shown in the inset.  The predicted yields at Daya Bay from {\sc Geant4} and {\sc Fluka} are also shown.  Experimental data is shown from Hertenberger~\cite{Hertenberger:1995ae}, Boehm~\cite{Boehm:2000ru}, Aberdeen Tunnel~\cite{Blyth:2015nha}, KamLAND~\cite{Abe:2009aa}, LVD~\cite{Aglietta:1999iw} with corrections from \cite{Mei:2005gm}, and Borexino~\cite{Bellini:2013pxa}.  The solid line shows the power-law fit to the global data set including Daya Bay.  The dashed line and dash-dotted lines show {\sc Fluka}-based predictions for the dependence of the neutron yield on muon energy from Wang {\it et al}~\cite{Wang:2001fq} and Kudryavtsey {\it et al}~\cite{Kudryavtsev:2003aua}.}
\end{figure*}

Comparisons of neutron yield measurements from different experiments are typically shown in terms of the average muon energy, despite the differences in muon energy distributions and angular distributions.  Daya Bay's measured values for the neutron yield for all three experimental halls are shown in Fig.~\ref{fig:Compare} as a function of average muon energy.  
%The points for Daya Bay EH1 and EH2 differ in energy by less than 1~GeV, and therefore overlap in the figure. 
The predicted yields at Daya Bay from {\sc Geant4} and {\sc Fluka} are also shown.  These results are compared with other experimental measurements over a wide range of average muon energies.  For the references that quote an average depth instead of average muon energy, the depth is converted to an average muon energy based on an average rock density~\cite{Mei:2005gm}.

Previous studies~\cite{Wang:2001fq,Kudryavtsev:2003aua,Araujo:2004rv,Mei:2005gm} have shown that the yield as a function of average muon energy can be described by a power-law,

\begin{align}
  Y_n = aE_{\mu}^b.
\end{align}
A power-law fit applied to the measurements shown in Fig.~\ref{fig:Compare} including the three points from Daya Bay yields coefficients $a = (4.0 \pm 0.6)\times 10^{-6}~\mu^{-1}~{\rm g}^{-1}~{\rm cm}^2$ and $b = 0.77\pm 0.03$.  
%chi2/dof = 5.143/7
 Not all the references quote an uncertainty in the depth or muon energy. Therefore, zero uncertainty in the average muon energy is assumed for all points included in the global fit. 

Daya Bay has the unique capability to measure neutron production at three different underground sites with essentially identical detectors.  A power-law fit applied to the three data points from Daya Bay (including the uncertainty in the average muon energy in the fit) yields coefficients $a = (7.2 \pm 3.8)\times 10^{-6}~\mu^{-1}~{\rm g}^{-1}~{\rm cm}^2$ and $b = 0.64\pm 0.12$.  
%chi2/dof = 0.009/1

In Fig.~\ref{fig:Compare}, the global fit is compared to {\sc Fluka}-based studies performed by Wang {\it et al}~\cite{Wang:2001fq} and Kudryavtsey {\it et al}~\cite{Kudryavtsev:2003aua}.  The {\sc Fluka} predictions from this study and Refs.~\cite{Wang:2001fq,Kudryavtsev:2003aua} are consistent with the measurement at EH3, but predict fewer neutrons for the shallower depths at EH1 and EH2.  Other measurements shown in Fig.~\ref{fig:Compare} show similar or even larger discrepancies with respect to the {\sc Fluka}-based predictions.  {\sc Geant4} has been shown to predict up to 30\% fewer neutrons than {\sc Fluka} at muon energies above 100~GeV~\cite{Araujo:2004rv}, which is consistent with the MC predictions in this analysis.

\section{Summary}
This paper presents the neutron yield in liquid scintillator measured at the three different experimental sites of the Daya Bay experiment, with different depths corresponding to different average muon energies.  These measurements are compared to the values predicted from {\sc Geant4} and {\sc Fluka} MC, revealing some possible discrepancies with the MC models.  A power-law fit of the dependence of the neutron yield on average muon energy is obtained by including the Daya Bay measurements with measurements from other experiments.

\begin{acknowledgments}
The Daya Bay Experiment is supported in part by
the Ministry of Science and Technology of China,
the United States Department of Energy,
the Chinese Academy of Sciences,
the CAS Center for Excellence in Particle Phsyics,
the National Natural Science Foundation of China,
the Guangdong provincial government,
the Shenzhen municipal government,
the China General Nuclear Power Group,
the Research Grants Council of the Hong Kong Special Administrative Region of China,
the Ministry of Education in Taiwan,
the U.S. National Science Foundation,
the Ministry of Education, Youth and Sports of the Czech Republic,
the Joint Institute of Nuclear Research in Dubna, Russia,
the NSFC-RFBR joint research program,
the National Commission for Scientific and Technological Research of Chile.
We acknowledge Yellow River Engineering Consulting Co., Ltd.\ and China Railway 15th Bureau Group Co., Ltd.\ for building the underground laboratory.
We are grateful for the ongoing cooperation from the China Guangdong Nuclear Power Group and China Light~\&~Power Company.

\end{acknowledgments}

\bibliography{Neutron}

\begin{thebibliography}{40}%
\makeatletter
\providecommand \@ifxundefined [1]{%
 \@ifx{#1\undefined}
}%
\providecommand \@ifnum [1]{%
 \ifnum #1\expandafter \@firstoftwo
 \else \expandafter \@secondoftwo
 \fi
}%
\providecommand \@ifx [1]{%
 \ifx #1\expandafter \@firstoftwo
 \else \expandafter \@secondoftwo
 \fi
}%
\providecommand \natexlab [1]{#1}%
\providecommand \enquote  [1]{``#1''}%
\providecommand \bibnamefont  [1]{#1}%
\providecommand \bibfnamefont [1]{#1}%
\providecommand \citenamefont [1]{#1}%
\providecommand \href@noop [0]{\@secondoftwo}%
\providecommand \href [0]{\begingroup \@sanitize@url \@href}%
\providecommand \@href[1]{\@@startlink{#1}\@@href}%
\providecommand \@@href[1]{\endgroup#1\@@endlink}%
\providecommand \@sanitize@url [0]{\catcode `\\12\catcode `\$12\catcode
  `\&12\catcode `\#12\catcode `\^12\catcode `\_12\catcode `\%12\relax}%
\providecommand \@@startlink[1]{}%
\providecommand \@@endlink[0]{}%
\providecommand \url  [0]{\begingroup\@sanitize@url \@url }%
\providecommand \@url [1]{\endgroup\@href {#1}{\urlprefix }}%
\providecommand \urlprefix  [0]{URL }%
\providecommand \Eprint [0]{\href }%
\providecommand \doibase [0]{http://dx.doi.org/}%
\providecommand \selectlanguage [0]{\@gobble}%
\providecommand \bibinfo  [0]{\@secondoftwo}%
\providecommand \bibfield  [0]{\@secondoftwo}%
\providecommand \translation [1]{[#1]}%
\providecommand \BibitemOpen [0]{}%
\providecommand \bibitemStop [0]{}%
\providecommand \bibitemNoStop [0]{.\EOS\space}%
\providecommand \EOS [0]{\spacefactor3000\relax}%
\providecommand \BibitemShut  [1]{\csname bibitem#1\endcsname}%
\let\auto@bib@innerbib\@empty
%</preamble>
\bibitem [{\citenamefont {Hagner}\ \emph {et~al.}(2000)\citenamefont {Hagner},
  \citenamefont {von Hentig}, \citenamefont {Heisinger}, \citenamefont
  {Oberauer}, \citenamefont {Sch{\"o}nert}, \citenamefont {von Feilitzsch},\
  and\ \citenamefont {Nolte}}]{Hagner:2000xb}%
  \BibitemOpen
  \bibfield  {author} {\bibinfo {author} {\bibfnamefont {T.}~\bibnamefont
  {Hagner}}, \bibinfo {author} {\bibfnamefont {R.}~\bibnamefont {von Hentig}},
  \bibinfo {author} {\bibfnamefont {B.}~\bibnamefont {Heisinger}}, \bibinfo
  {author} {\bibfnamefont {L.}~\bibnamefont {Oberauer}}, \bibinfo {author}
  {\bibfnamefont {S.}~\bibnamefont {Sch{\"o}nert}}, \bibinfo {author}
  {\bibfnamefont {F.}~\bibnamefont {von Feilitzsch}}, \ and\ \bibinfo {author}
  {\bibfnamefont {E.}~\bibnamefont {Nolte}},\ }\href {\doibase
  10.1016/S0927-6505(00)00103-1} {\bibfield  {journal} {\bibinfo  {journal}
  {Astropart.Phys.}\ }\textbf {\bibinfo {volume} {14}},\ \bibinfo {pages} {33}
  (\bibinfo {year} {2000})}\BibitemShut {NoStop}%
%%CITATION = APHYE,14,33;%%
\bibitem [{\citenamefont {Aglietta}\ \emph {et~al.}(shed)\citenamefont
  {Aglietta} \emph {et~al.}}]{Aglietta:1999iw}%
  \BibitemOpen
  \bibfield  {author} {\bibinfo {author} {\bibfnamefont {M.}~\bibnamefont
  {Aglietta}} \emph {et~al.} (\bibinfo {collaboration} {LVD Collaboration}),\
  }in\ \href {http://krusty.physics.utah.edu/~icrc1999/root/vol2/h3_1_15.pdf}
  {\emph {\bibinfo {booktitle} {{Proceedings, 26th International Cosmic Ray
  Conference, 1999, Salt Lake City: Invited, Rapporteur, and Highlight
  Papers}}}}\ (\bibinfo {year} {unpublished})\ \Eprint
  {http://arxiv.org/abs/hep-ex/9905047} {arXiv:hep-ex/9905047 [hep-ex]}
  \BibitemShut {NoStop}%
%%CITATION = HEP-EX/9905047;%%
\bibitem [{\citenamefont {Bellini}\ \emph {et~al.}(2013)\citenamefont {Bellini}
  \emph {et~al.}}]{Bellini:2013pxa}%
  \BibitemOpen
  \bibfield  {author} {\bibinfo {author} {\bibfnamefont {G.}~\bibnamefont
  {Bellini}} \emph {et~al.} (\bibinfo {collaboration} {Borexino
  Collaboration}),\ }\href {\doibase 10.1088/1475-7516/2013/08/049} {\bibfield
  {journal} {\bibinfo  {journal} {JCAP}\ }\textbf {\bibinfo {volume} {08}},\
  \bibinfo {pages} {049} (\bibinfo {year} {2013})},\ \Eprint
  {http://arxiv.org/abs/1304.7381} {arXiv:1304.7381 [physics.ins-det]}
  \BibitemShut {NoStop}%
%%CITATION = ARXIV:1304.7381;%%
\bibitem [{\citenamefont {Abe}\ \emph {et~al.}(2010)\citenamefont {Abe} \emph
  {et~al.}}]{Abe:2009aa}%
  \BibitemOpen
  \bibfield  {author} {\bibinfo {author} {\bibfnamefont {S.}~\bibnamefont
  {Abe}} \emph {et~al.} (\bibinfo {collaboration} {KamLAND Collaboration}),\
  }\href {\doibase 10.1103/PhysRevC.81.025807} {\bibfield  {journal} {\bibinfo
  {journal} {Phys.Rev.}\ }\textbf {\bibinfo {volume} {C81}},\ \bibinfo {pages}
  {025807} (\bibinfo {year} {2010})},\ \Eprint {http://arxiv.org/abs/0907.0066}
  {arXiv:0907.0066 [hep-ex]} \BibitemShut {NoStop}%
%%CITATION = ARXIV:0907.0066;%%
\bibitem [{\citenamefont {Aglietta}\ \emph {et~al.}(1989)\citenamefont
  {Aglietta} \emph {et~al.}}]{Aglietta:1989xn}%
  \BibitemOpen
  \bibfield  {author} {\bibinfo {author} {\bibfnamefont {M.}~\bibnamefont
  {Aglietta}} \emph {et~al.},\ }\href {\doibase 10.1007/BF02525079} {\bibfield
  {journal} {\bibinfo  {journal} {Il Nuovo Cim.}\ }\textbf {\bibinfo {volume}
  {C12}},\ \bibinfo {pages} {467} (\bibinfo {year} {1989})}\BibitemShut
  {NoStop}%
%%CITATION = NUCIA,C12,467;%%
\bibitem [{\citenamefont {Hertenberger}\ \emph {et~al.}(1995)\citenamefont
  {Hertenberger}, \citenamefont {Chen},\ and\ \citenamefont
  {Dougherty}}]{Hertenberger:1995ae}%
  \BibitemOpen
  \bibfield  {author} {\bibinfo {author} {\bibfnamefont {R.}~\bibnamefont
  {Hertenberger}}, \bibinfo {author} {\bibfnamefont {M.}~\bibnamefont {Chen}},
  \ and\ \bibinfo {author} {\bibfnamefont {B.~L.}\ \bibnamefont {Dougherty}},\
  }\href {\doibase 10.1103/PhysRevC.52.3449} {\bibfield  {journal} {\bibinfo
  {journal} {Phys. Rev.}\ }\textbf {\bibinfo {volume} {C52}},\ \bibinfo {pages}
  {3449} (\bibinfo {year} {1995})}\BibitemShut {NoStop}%
%%CITATION = PHRVA,C52,3449;%%
\bibitem [{\citenamefont {Bezrukov}\ \emph {et~al.}(1973)\citenamefont
  {Bezrukov} \emph {et~al.}}]{Bezrukov}%
  \BibitemOpen
  \bibfield  {author} {\bibinfo {author} {\bibfnamefont {L.~B.}\ \bibnamefont
  {Bezrukov}} \emph {et~al.},\ }\href@noop {} {\bibfield  {journal} {\bibinfo
  {journal} {Sov. J. Nucl. Phys.}\ }\textbf {\bibinfo {volume} {17}},\ \bibinfo
  {pages} {51} (\bibinfo {year} {1973})},\ \bibinfo {note} {[Yad.
  Fiz.17,98(1973)]}\BibitemShut {NoStop}%
\bibitem [{\citenamefont {Boehm}\ \emph {et~al.}(2000)\citenamefont {Boehm}
  \emph {et~al.}}]{Boehm:2000ru}%
  \BibitemOpen
  \bibfield  {author} {\bibinfo {author} {\bibfnamefont {F.}~\bibnamefont
  {Boehm}} \emph {et~al.},\ }\href {\doibase 10.1103/PhysRevD.62.092005}
  {\bibfield  {journal} {\bibinfo  {journal} {Phys. Rev.}\ }\textbf {\bibinfo
  {volume} {D62}},\ \bibinfo {pages} {092005} (\bibinfo {year} {2000})},\
  \Eprint {http://arxiv.org/abs/hep-ex/0006014} {arXiv:hep-ex/0006014 [hep-ex]}
  \BibitemShut {NoStop}%
%%CITATION = HEP-EX/0006014;%%
\bibitem [{\citenamefont {Enikeev}\ \emph {et~al.}(1987)\citenamefont {Enikeev}
  \emph {et~al.}}]{Enikeev}%
  \BibitemOpen
  \bibfield  {author} {\bibinfo {author} {\bibfnamefont {R.~I.}\ \bibnamefont
  {Enikeev}} \emph {et~al.},\ }\href@noop {} {\bibfield  {journal} {\bibinfo
  {journal} {Sov. J. Nucl. Phys.}\ }\textbf {\bibinfo {volume} {46}},\ \bibinfo
  {pages} {883} (\bibinfo {year} {1987})},\ \bibinfo {note} {[Yad.
  Fiz.46,1492(1987)]}\BibitemShut {NoStop}%
\bibitem [{\citenamefont {Blyth}\ \emph {et~al.}(2016)\citenamefont {Blyth}
  \emph {et~al.}}]{Blyth:2015nha}%
  \BibitemOpen
  \bibfield  {author} {\bibinfo {author} {\bibfnamefont {S.~C.}\ \bibnamefont
  {Blyth}} \emph {et~al.} (\bibinfo {collaboration} {Aberdeen Tunnel Experiment
  Collaboration}),\ }\href {\doibase 10.1103/PhysRevD.93.072005} {\bibfield
  {journal} {\bibinfo  {journal} {Phys. Rev.}\ }\textbf {\bibinfo {volume}
  {D93}},\ \bibinfo {pages} {072005} (\bibinfo {year} {2016})},\ \Eprint
  {http://arxiv.org/abs/1509.09038} {arXiv:1509.09038 [physics.ins-det]}
  \BibitemShut {NoStop}%
%%CITATION = ARXIV:1509.09038;%%
\bibitem [{\citenamefont {Araujo}\ \emph {et~al.}(2008)\citenamefont {Araujo}
  \emph {et~al.}}]{Araujo:2008ze}%
  \BibitemOpen
  \bibfield  {author} {\bibinfo {author} {\bibfnamefont {H.~M.}\ \bibnamefont
  {Araujo}} \emph {et~al.},\ }\href {\doibase
  10.1016/j.astropartphys.2008.05.004} {\bibfield  {journal} {\bibinfo
  {journal} {Astropart. Phys.}\ }\textbf {\bibinfo {volume} {29}},\ \bibinfo
  {pages} {471} (\bibinfo {year} {2008})},\ \Eprint
  {http://arxiv.org/abs/0805.3110} {arXiv:0805.3110 [hep-ex]} \BibitemShut
  {NoStop}%
%%CITATION = ARXIV:0805.3110;%%
\bibitem [{\citenamefont {Reichhart}\ \emph {et~al.}(2013)\citenamefont
  {Reichhart} \emph {et~al.}}]{Reichhart:2013xkd}%
  \BibitemOpen
  \bibfield  {author} {\bibinfo {author} {\bibfnamefont {L.}~\bibnamefont
  {Reichhart}} \emph {et~al.},\ }\href {\doibase
  10.1016/j.astropartphys.2013.06.002} {\bibfield  {journal} {\bibinfo
  {journal} {Astropart. Phys.}\ }\textbf {\bibinfo {volume} {47}},\ \bibinfo
  {pages} {67} (\bibinfo {year} {2013})},\ \Eprint
  {http://arxiv.org/abs/1302.4275} {arXiv:1302.4275 [physics.ins-det]}
  \BibitemShut {NoStop}%
%%CITATION = ARXIV:1302.4275;%%
\bibitem [{\citenamefont {Guo}\ \emph {et~al.}(2007)\citenamefont {Guo} \emph
  {et~al.}}]{Guo:2007ug}%
  \BibitemOpen
  \bibfield  {author} {\bibinfo {author} {\bibfnamefont {X.}~\bibnamefont
  {Guo}} \emph {et~al.} (\bibinfo {collaboration} {Daya Bay Collaboration}),\
  }\href@noop {} {\  (\bibinfo {year} {2007})},\ \Eprint
  {http://arxiv.org/abs/hep-ex/0701029} {arXiv:hep-ex/0701029 [hep-ex]}
  \BibitemShut {NoStop}%
%%CITATION = HEP-EX/0701029;%%
\bibitem [{\citenamefont {An}\ \emph {et~al.}(2012{\natexlab{a}})\citenamefont
  {An} \emph {et~al.}}]{An:2012eh}%
  \BibitemOpen
  \bibfield  {author} {\bibinfo {author} {\bibfnamefont {F.~P.}\ \bibnamefont
  {An}} \emph {et~al.} (\bibinfo {collaboration} {Daya Bay Collaboration}),\
  }\href {\doibase 10.1103/PhysRevLett.108.171803} {\bibfield  {journal}
  {\bibinfo  {journal} {Phys.Rev.Lett.}\ }\textbf {\bibinfo {volume} {108}},\
  \bibinfo {pages} {171803} (\bibinfo {year} {2012}{\natexlab{a}})},\ \Eprint
  {http://arxiv.org/abs/1203.1669} {arXiv:1203.1669 [hep-ex]} \BibitemShut
  {NoStop}%
%%CITATION = ARXIV:1203.1669;%%
\bibitem [{\citenamefont {An}\ \emph {et~al.}(2013)\citenamefont {An} \emph
  {et~al.}}]{An:2013uza}%
  \BibitemOpen
  \bibfield  {author} {\bibinfo {author} {\bibfnamefont {F.~P.}\ \bibnamefont
  {An}} \emph {et~al.} (\bibinfo {collaboration} {Daya Bay Collaboration}),\
  }\href {\doibase 10.1088/1674-1137/37/1/011001} {\bibfield  {journal}
  {\bibinfo  {journal} {Chin.Phys.}\ }\textbf {\bibinfo {volume} {C37}},\
  \bibinfo {pages} {011001} (\bibinfo {year} {2013})},\ \Eprint
  {http://arxiv.org/abs/1210.6327} {arXiv:1210.6327 [hep-ex]} \BibitemShut
  {NoStop}%
%%CITATION = ARXIV:1210.6327;%%
\bibitem [{\citenamefont {An}\ \emph {et~al.}(2014)\citenamefont {An} \emph
  {et~al.}}]{An:2013zwz}%
  \BibitemOpen
  \bibfield  {author} {\bibinfo {author} {\bibfnamefont {F.~P.}\ \bibnamefont
  {An}} \emph {et~al.} (\bibinfo {collaboration} {Daya Bay Collaboration}),\
  }\href {\doibase 10.1103/PhysRevLett.112.061801} {\bibfield  {journal}
  {\bibinfo  {journal} {Phys.Rev.Lett.}\ }\textbf {\bibinfo {volume} {112}},\
  \bibinfo {pages} {061801} (\bibinfo {year} {2014})},\ \Eprint
  {http://arxiv.org/abs/1310.6732} {arXiv:1310.6732 [hep-ex]} \BibitemShut
  {NoStop}%
%%CITATION = ARXIV:1310.6732;%%
\bibitem [{\citenamefont {An}\ \emph {et~al.}(2015{\natexlab{a}})\citenamefont
  {An} \emph {et~al.}}]{An:2015rpe}%
  \BibitemOpen
  \bibfield  {author} {\bibinfo {author} {\bibfnamefont {F.~P.}\ \bibnamefont
  {An}} \emph {et~al.} (\bibinfo {collaboration} {Daya Bay Collaboration}),\
  }\href {\doibase 10.1103/PhysRevLett.115.111802} {\bibfield  {journal}
  {\bibinfo  {journal} {Phys. Rev. Lett.}\ }\textbf {\bibinfo {volume} {115}},\
  \bibinfo {pages} {111802} (\bibinfo {year} {2015}{\natexlab{a}})},\ \Eprint
  {http://arxiv.org/abs/1505.03456} {arXiv:1505.03456 [hep-ex]} \BibitemShut
  {NoStop}%
%%CITATION = ARXIV:1505.03456;%%
\bibitem [{\citenamefont {An}\ \emph {et~al.}(2017{\natexlab{a}})\citenamefont
  {An} \emph {et~al.}}]{An:2016ses}%
  \BibitemOpen
  \bibfield  {author} {\bibinfo {author} {\bibfnamefont {F.~P.}\ \bibnamefont
  {An}} \emph {et~al.} (\bibinfo {collaboration} {Daya Bay Collaboration}),\
  }\href {\doibase 10.1103/PhysRevD.95.072006} {\bibfield  {journal} {\bibinfo
  {journal} {Phys. Rev.}\ }\textbf {\bibinfo {volume} {D95}},\ \bibinfo {pages}
  {072006} (\bibinfo {year} {2017}{\natexlab{a}})},\ \Eprint
  {http://arxiv.org/abs/1610.04802} {arXiv:1610.04802 [hep-ex]} \BibitemShut
  {NoStop}%
%%CITATION = ARXIV:1610.04802;%%
\bibitem [{\citenamefont {An}\ \emph {et~al.}(2012{\natexlab{b}})\citenamefont
  {An} \emph {et~al.}}]{DayaBay:2012aa}%
  \BibitemOpen
  \bibfield  {author} {\bibinfo {author} {\bibfnamefont {F.~P.}\ \bibnamefont
  {An}} \emph {et~al.} (\bibinfo {collaboration} {Daya Bay Collaboration}),\
  }\href {\doibase 10.1016/j.nima.2012.05.030} {\bibfield  {journal} {\bibinfo
  {journal} {Nucl.Instrum.Meth.}\ }\textbf {\bibinfo {volume} {A685}},\
  \bibinfo {pages} {78} (\bibinfo {year} {2012}{\natexlab{b}})},\ \Eprint
  {http://arxiv.org/abs/1202.6181} {arXiv:1202.6181 [physics.ins-det]}
  \BibitemShut {NoStop}%
%%CITATION = ARXIV:1202.6181;%%
\bibitem [{\citenamefont {An}\ \emph {et~al.}(2016{\natexlab{a}})\citenamefont
  {An} \emph {et~al.}}]{An:2015qga}%
  \BibitemOpen
  \bibfield  {author} {\bibinfo {author} {\bibfnamefont {F.~P.}\ \bibnamefont
  {An}} \emph {et~al.} (\bibinfo {collaboration} {Daya Bay Collaboration}),\
  }\href {\doibase 10.1016/j.nima.2015.11.144} {\bibfield  {journal} {\bibinfo
  {journal} {Nucl. Instrum. Meth.}\ }\textbf {\bibinfo {volume} {A811}},\
  \bibinfo {pages} {133} (\bibinfo {year} {2016}{\natexlab{a}})},\ \Eprint
  {http://arxiv.org/abs/1508.03943} {arXiv:1508.03943 [physics.ins-det]}
  \BibitemShut {NoStop}%
%%CITATION = ARXIV:1508.03943;%%
\bibitem [{\citenamefont {An}\ \emph {et~al.}(2015{\natexlab{b}})\citenamefont
  {An} \emph {et~al.}}]{Dayabay:2014vka}%
  \BibitemOpen
  \bibfield  {author} {\bibinfo {author} {\bibfnamefont {F.~P.}\ \bibnamefont
  {An}} \emph {et~al.} (\bibinfo {collaboration} {Daya Bay Collaboration}),\
  }\href {\doibase 10.1016/j.nima.2014.09.070} {\bibfield  {journal} {\bibinfo
  {journal} {Nucl.Instrum.Meth.}\ }\textbf {\bibinfo {volume} {A773}},\
  \bibinfo {pages} {8} (\bibinfo {year} {2015}{\natexlab{b}})},\ \Eprint
  {http://arxiv.org/abs/1407.0275} {arXiv:1407.0275 [physics.ins-det]}
  \BibitemShut {NoStop}%
%%CITATION = ARXIV:1407.0275;%%
\bibitem [{\citenamefont {Wilhelmi}\ \emph {et~al.}(2015)\citenamefont
  {Wilhelmi} \emph {et~al.}}]{Wilhelmi:2014irz}%
  \BibitemOpen
  \bibfield  {author} {\bibinfo {author} {\bibfnamefont {J.}~\bibnamefont
  {Wilhelmi}} \emph {et~al.},\ }\href@noop {} {\bibfield  {journal} {\bibinfo
  {journal} {Journal of Water Process Engineering}\ }\textbf {\bibinfo {volume}
  {5}},\ \bibinfo {pages} {127} (\bibinfo {year} {2015})},\ \Eprint
  {http://arxiv.org/abs/1408.1302} {arXiv:1408.1302 [physics.ins-det]}
  \BibitemShut {NoStop}%
%%CITATION = ARXIV:1408.1302;%%
\bibitem [{\citenamefont {Gaisser}\ \emph {et~al.}()\citenamefont {Gaisser},
  \citenamefont {Engel},\ and\ \citenamefont {Resconi}}]{Gaisser1}%
  \BibitemOpen
  \bibfield  {author} {\bibinfo {author} {\bibfnamefont {T.~K.}\ \bibnamefont
  {Gaisser}}, \bibinfo {author} {\bibfnamefont {R.}~\bibnamefont {Engel}}, \
  and\ \bibinfo {author} {\bibfnamefont {E.}~\bibnamefont {Resconi}},\ }\href
  {http://www.cambridge.org/de/academic/subjects/physics/cosmology-relativity-and-gravitation/cosmic-rays-and-particle-physics-2nd-edition?format=HB}
  {\emph {\bibinfo {title} {{Cosmic Rays and Particle Physics}}}}\ (\bibinfo
  {publisher} {Cambridge University Press, Cambridge, 2016})\BibitemShut
  {NoStop}%
%%CITATION = INSPIRE-1419789;%%
\bibitem [{\citenamefont {Eidelman}\ \emph {et~al.}(2004)\citenamefont
  {Eidelman} \emph {et~al.}}]{Eidelman:2004wy}%
  \BibitemOpen
  \bibfield  {author} {\bibinfo {author} {\bibfnamefont {S.}~\bibnamefont
  {Eidelman}} \emph {et~al.} (\bibinfo {collaboration} {Particle Data Group}),\
  }\href {\doibase 10.1016/j.physletb.2004.06.001} {\bibfield  {journal}
  {\bibinfo  {journal} {Phys. Lett.}\ }\textbf {\bibinfo {volume} {B592}},\
  \bibinfo {pages} {1} (\bibinfo {year} {2004})}\BibitemShut {NoStop}%
%%CITATION = PHLTA,B592,1;%%
\bibitem [{\citenamefont {Guan}\ \emph {et~al.}(2006)\citenamefont {Guan} \emph
  {et~al.}}]{Guan}%
  \BibitemOpen
  \bibfield  {author} {\bibinfo {author} {\bibfnamefont {M.}~\bibnamefont
  {Guan}} \emph {et~al.},\ }\href@noop {} {\  (\bibinfo {year} {2006})},\
  \bibinfo {note} {{Lawrence Berkeley National Laboratory Paper
  LBNL-4262E}}\BibitemShut {NoStop}%
\bibitem [{\citenamefont {Antonioli}\ \emph {et~al.}(1997)\citenamefont
  {Antonioli}, \citenamefont {Ghetti}, \citenamefont {Korolkova}, \citenamefont
  {Kudryavtsev},\ and\ \citenamefont {Sartorelli}}]{Antonioli:1997qw}%
  \BibitemOpen
  \bibfield  {author} {\bibinfo {author} {\bibfnamefont {P.}~\bibnamefont
  {Antonioli}}, \bibinfo {author} {\bibfnamefont {C.}~\bibnamefont {Ghetti}},
  \bibinfo {author} {\bibfnamefont {E.~V.}\ \bibnamefont {Korolkova}}, \bibinfo
  {author} {\bibfnamefont {V.~A.}\ \bibnamefont {Kudryavtsev}}, \ and\ \bibinfo
  {author} {\bibfnamefont {G.}~\bibnamefont {Sartorelli}},\ }\href {\doibase
  10.1016/S0927-6505(97)00035-2} {\bibfield  {journal} {\bibinfo  {journal}
  {Astropart. Phys.}\ }\textbf {\bibinfo {volume} {7}},\ \bibinfo {pages} {357}
  (\bibinfo {year} {1997})},\ \Eprint {http://arxiv.org/abs/hep-ph/9705408}
  {arXiv:hep-ph/9705408 [hep-ph]} \BibitemShut {NoStop}%
%%CITATION = HEP-PH/9705408;%%
\bibitem [{\citenamefont {Kudryavtsev}(2009)}]{Kudryavtsev:2008qh}%
  \BibitemOpen
  \bibfield  {author} {\bibinfo {author} {\bibfnamefont {V.~A.}\ \bibnamefont
  {Kudryavtsev}},\ }\href {\doibase 10.1016/j.cpc.2008.10.013} {\bibfield
  {journal} {\bibinfo  {journal} {Comput. Phys. Commun.}\ }\textbf {\bibinfo
  {volume} {180}},\ \bibinfo {pages} {339} (\bibinfo {year} {2009})},\ \Eprint
  {http://arxiv.org/abs/0810.4635} {arXiv:0810.4635 [physics.comp-ph]}
  \BibitemShut {NoStop}%
%%CITATION = ARXIV:0810.4635;%%
\bibitem [{\citenamefont {Agostinelli}\ \emph {et~al.}(2003)\citenamefont
  {Agostinelli} \emph {et~al.}}]{Agostinelli:2002hh}%
  \BibitemOpen
  \bibfield  {author} {\bibinfo {author} {\bibfnamefont {S.}~\bibnamefont
  {Agostinelli}} \emph {et~al.} (\bibinfo {collaboration} {GEANT4
  Collaboration}),\ }\href {\doibase 10.1016/S0168-9002(03)01368-8} {\bibfield
  {journal} {\bibinfo  {journal} {Nucl. Instrum. Meth.}\ }\textbf {\bibinfo
  {volume} {A506}},\ \bibinfo {pages} {250} (\bibinfo {year}
  {2003})}\BibitemShut {NoStop}%
%%CITATION = NUIMA,A506,250;%%
\bibitem [{\citenamefont {Allison}\ \emph {et~al.}(2006)\citenamefont {Allison}
  \emph {et~al.}}]{Allison:2006ve}%
  \BibitemOpen
  \bibfield  {author} {\bibinfo {author} {\bibfnamefont {J.}~\bibnamefont
  {Allison}} \emph {et~al.},\ }\href {\doibase 10.1109/TNS.2006.869826}
  {\bibfield  {journal} {\bibinfo  {journal} {IEEE Trans. Nucl. Sci.}\ }\textbf
  {\bibinfo {volume} {53}},\ \bibinfo {pages} {270} (\bibinfo {year}
  {2006})}\BibitemShut {NoStop}%
%%CITATION = IETNA,53,270;%%
\bibitem [{\citenamefont {Boehlen}\ \emph {et~al.}(2014)\citenamefont {Boehlen}
  \emph {et~al.}}]{fluka1}%
  \BibitemOpen
  \bibfield  {author} {\bibinfo {author} {\bibfnamefont {T.~T.}\ \bibnamefont
  {Boehlen}} \emph {et~al.},\ }\href@noop {} {\bibfield  {journal} {\bibinfo
  {journal} {Nuclear Data Sheets}\ }\textbf {\bibinfo {volume} {120}},\
  \bibinfo {pages} {211} (\bibinfo {year} {2014})}\BibitemShut {NoStop}%
\bibitem [{\citenamefont {Ferrari}\ \emph {et~al.}(2005)\citenamefont
  {Ferrari}, \citenamefont {Sala}, \citenamefont {Fasso},\ and\ \citenamefont
  {Ranft}}]{Ferrari:2005zk}%
  \BibitemOpen
  \bibfield  {author} {\bibinfo {author} {\bibfnamefont {A.}~\bibnamefont
  {Ferrari}}, \bibinfo {author} {\bibfnamefont {P.~R.}\ \bibnamefont {Sala}},
  \bibinfo {author} {\bibfnamefont {A.}~\bibnamefont {Fasso}}, \ and\ \bibinfo
  {author} {\bibfnamefont {J.}~\bibnamefont {Ranft}},\ }\href@noop {} {\
  (\bibinfo {year} {2005})},\ \bibinfo {note} {{Report Nos. CERN-2005-010,
  SLAC-R-773, INFN-TC-05-11, http://cds.cern.ch/record/898301}}\BibitemShut
  {NoStop}%
%%CITATION = CERN-2005-010;%%
\bibitem [{\citenamefont {Wang}\ \emph {et~al.}(2001)\citenamefont {Wang},
  \citenamefont {Balic}, \citenamefont {Gratta}, \citenamefont {Fasso},
  \citenamefont {Roesler},\ and\ \citenamefont {Ferrari}}]{Wang:2001fq}%
  \BibitemOpen
  \bibfield  {author} {\bibinfo {author} {\bibfnamefont {Y.~F.}\ \bibnamefont
  {Wang}}, \bibinfo {author} {\bibfnamefont {V.}~\bibnamefont {Balic}},
  \bibinfo {author} {\bibfnamefont {G.}~\bibnamefont {Gratta}}, \bibinfo
  {author} {\bibfnamefont {A.}~\bibnamefont {Fasso}}, \bibinfo {author}
  {\bibfnamefont {S.}~\bibnamefont {Roesler}}, \ and\ \bibinfo {author}
  {\bibfnamefont {A.}~\bibnamefont {Ferrari}},\ }\href {\doibase
  10.1103/PhysRevD.64.013012} {\bibfield  {journal} {\bibinfo  {journal} {Phys.
  Rev.}\ }\textbf {\bibinfo {volume} {D64}},\ \bibinfo {pages} {013012}
  (\bibinfo {year} {2001})},\ \Eprint {http://arxiv.org/abs/hep-ex/0101049}
  {arXiv:hep-ex/0101049 [hep-ex]} \BibitemShut {NoStop}%
%%CITATION = HEP-EX/0101049;%%
\bibitem [{\citenamefont {Kudryavtsev}\ \emph {et~al.}(2003)\citenamefont
  {Kudryavtsev}, \citenamefont {Spooner},\ and\ \citenamefont
  {McMillan}}]{Kudryavtsev:2003aua}%
  \BibitemOpen
  \bibfield  {author} {\bibinfo {author} {\bibfnamefont {V.~A.}\ \bibnamefont
  {Kudryavtsev}}, \bibinfo {author} {\bibfnamefont {N.~J.~C.}\ \bibnamefont
  {Spooner}}, \ and\ \bibinfo {author} {\bibfnamefont {J.~E.}\ \bibnamefont
  {McMillan}},\ }\href {\doibase 10.1016/S0168-9002(03)00983-5} {\bibfield
  {journal} {\bibinfo  {journal} {Nucl. Instrum. Meth.}\ }\textbf {\bibinfo
  {volume} {A505}},\ \bibinfo {pages} {688} (\bibinfo {year} {2003})},\ \Eprint
  {http://arxiv.org/abs/hep-ex/0303007} {arXiv:hep-ex/0303007 [hep-ex]}
  \BibitemShut {NoStop}%
%%CITATION = HEP-EX/0303007;%%
\bibitem [{\citenamefont {Araujo}\ \emph {et~al.}(2005)\citenamefont {Araujo},
  \citenamefont {Kudryavtsev}, \citenamefont {Spooner},\ and\ \citenamefont
  {Sumner}}]{Araujo:2004rv}%
  \BibitemOpen
  \bibfield  {author} {\bibinfo {author} {\bibfnamefont {H.~M.}\ \bibnamefont
  {Araujo}}, \bibinfo {author} {\bibfnamefont {V.~A.}\ \bibnamefont
  {Kudryavtsev}}, \bibinfo {author} {\bibfnamefont {N.~J.~C.}\ \bibnamefont
  {Spooner}}, \ and\ \bibinfo {author} {\bibfnamefont {T.~J.}\ \bibnamefont
  {Sumner}},\ }\href {\doibase 10.1016/j.nima.2005.02.004} {\bibfield
  {journal} {\bibinfo  {journal} {Nucl. Instrum. Meth.}\ }\textbf {\bibinfo
  {volume} {A545}},\ \bibinfo {pages} {398} (\bibinfo {year} {2005})},\ \Eprint
  {http://arxiv.org/abs/hep-ex/0411026} {arXiv:hep-ex/0411026 [hep-ex]}
  \BibitemShut {NoStop}%
%%CITATION = HEP-EX/0411026;%%
\bibitem [{\citenamefont {Mei}\ and\ \citenamefont {Hime}(2006)}]{Mei:2005gm}%
  \BibitemOpen
  \bibfield  {author} {\bibinfo {author} {\bibfnamefont {D.}~\bibnamefont
  {Mei}}\ and\ \bibinfo {author} {\bibfnamefont {A.}~\bibnamefont {Hime}},\
  }\href {\doibase 10.1103/PhysRevD.73.053004} {\bibfield  {journal} {\bibinfo
  {journal} {Phys. Rev.}\ }\textbf {\bibinfo {volume} {D73}},\ \bibinfo {pages}
  {053004} (\bibinfo {year} {2006})},\ \Eprint
  {http://arxiv.org/abs/astro-ph/0512125} {arXiv:astro-ph/0512125 [astro-ph]}
  \BibitemShut {NoStop}%
%%CITATION = ASTRO-PH/0512125;%%
\bibitem [{\citenamefont {Empl}\ \emph {et~al.}(2014)\citenamefont {Empl},
  \citenamefont {Hungerford}, \citenamefont {Jasem},\ and\ \citenamefont
  {Mosteiro}}]{Empl:2014zea}%
  \BibitemOpen
  \bibfield  {author} {\bibinfo {author} {\bibfnamefont {A.}~\bibnamefont
  {Empl}}, \bibinfo {author} {\bibfnamefont {E.~V.}\ \bibnamefont
  {Hungerford}}, \bibinfo {author} {\bibfnamefont {R.}~\bibnamefont {Jasem}}, \
  and\ \bibinfo {author} {\bibfnamefont {P.}~\bibnamefont {Mosteiro}},\ }\href
  {\doibase 10.1088/1475-7516/2014/08/064} {\bibfield  {journal} {\bibinfo
  {journal} {JCAP}\ }\textbf {\bibinfo {volume} {1408}},\ \bibinfo {pages}
  {064} (\bibinfo {year} {2014})},\ \Eprint {http://arxiv.org/abs/1406.6081}
  {arXiv:1406.6081 [astro-ph.IM]} \BibitemShut {NoStop}%
%%CITATION = ARXIV:1406.6081;%%
\bibitem [{\citenamefont {Formaggio}\ and\ \citenamefont
  {Martoff}(2004)}]{Formaggio:2004ge}%
  \BibitemOpen
  \bibfield  {author} {\bibinfo {author} {\bibfnamefont {J.~A.}\ \bibnamefont
  {Formaggio}}\ and\ \bibinfo {author} {\bibfnamefont {C.~J.}\ \bibnamefont
  {Martoff}},\ }\href {\doibase 10.1146/annurev.nucl.54.070103.181248}
  {\bibfield  {journal} {\bibinfo  {journal} {Ann. Rev. Nucl. Part. Sci.}\
  }\textbf {\bibinfo {volume} {54}},\ \bibinfo {pages} {361} (\bibinfo {year}
  {2004})}\BibitemShut {NoStop}%
%%CITATION = ARNUA,54,361;%%
\bibitem [{\citenamefont {Ferrari}\ and\ \citenamefont
  {Sala}(shed)}]{Ferrari:1993xr}%
  \BibitemOpen
  \bibfield  {author} {\bibinfo {author} {\bibfnamefont {A.}~\bibnamefont
  {Ferrari}}\ and\ \bibinfo {author} {\bibfnamefont {P.~R.}\ \bibnamefont
  {Sala}},\ }in\ \href@noop {} {\emph {\bibinfo {booktitle} {{International
  Conference on Monte Carlo Simulation in High-Energy and Nuclear Physics - MC
  93 Tallahassee, Florida, 1993}}}}\ (\bibinfo {year}
  {unpublished})\BibitemShut {NoStop}%
%%CITATION = INSPIRE-367781;%%
\bibitem [{\citenamefont {An}\ \emph {et~al.}(2016{\natexlab{b}})\citenamefont
  {An} \emph {et~al.}}]{An:2015nua}%
  \BibitemOpen
  \bibfield  {author} {\bibinfo {author} {\bibfnamefont {F.~P.}\ \bibnamefont
  {An}} \emph {et~al.} (\bibinfo {collaboration} {Daya Bay Collaboration}),\
  }\href {\doibase 10.1103/PhysRevLett.116.061801} {\bibfield  {journal}
  {\bibinfo  {journal} {Phys. Rev. Lett.}\ }\textbf {\bibinfo {volume} {116}},\
  \bibinfo {pages} {061801} (\bibinfo {year} {2016}{\natexlab{b}})},\ \Eprint
  {http://arxiv.org/abs/1508.04233} {arXiv:1508.04233 [hep-ex]} \BibitemShut
  {NoStop}%
%%CITATION = ARXIV:1508.04233;%%
\bibitem [{\citenamefont {An}\ \emph {et~al.}(2017{\natexlab{b}})\citenamefont
  {An} \emph {et~al.}}]{An:2016srz}%
  \BibitemOpen
  \bibfield  {author} {\bibinfo {author} {\bibfnamefont {F.~P.}\ \bibnamefont
  {An}} \emph {et~al.} (\bibinfo {collaboration} {Daya Bay Collaboration}),\
  }\href {\doibase 10.1088/1674-1137/41/1/013002} {\bibfield  {journal}
  {\bibinfo  {journal} {Chin. Phys.}\ }\textbf {\bibinfo {volume} {C41}},\
  \bibinfo {pages} {013002} (\bibinfo {year} {2017}{\natexlab{b}})},\ \Eprint
  {http://arxiv.org/abs/1607.05378} {arXiv:1607.05378 [hep-ex]} \BibitemShut
  {NoStop}%
%%CITATION = ARXIV:1607.05378;%%
\end{thebibliography}%

\end{document}